\documentclass[12pt]{article}
\usepackage{amsmath,amsfonts, feynmp, epsf}
\usepackage{array}
\usepackage{makecell}
\newcolumntype{x}[1]{>{\centering\arraybackslash}p{#1}}

\usepackage{tikz}
\newcommand\diag[4]{
  \multicolumn{1}{p{#2}|}{\hskip-\tabcolsep
  $\vcenter{\begin{tikzpicture}[baseline=0,anchor=south west,inner sep=#1]
  \path[use as bounding box] (0,0) rectangle (#2+2\tabcolsep,\baselineskip);
  \node[minimum width={#2+2\tabcolsep},minimum height=\baselineskip+\extrarowheight] (box) {};
  \draw (box.north west) -- (box.south east);
    \path[use as bounding box] (0,0) rectangle (#2+2\tabcolsep,\baselineskip);
  \node[minimum width={#2+2\tabcolsep},minimum height=\baselineskip+\extrarowheight] (box) {};
  \draw (box.north west) -- (box.south west);
  \node[anchor=south west] at (box.south west) {#3};
  \node[anchor=north east] at (box.north east) {#4};
 \end{tikzpicture}}$\hskip-\tabcolsep}}

\newcolumntype{x}[1]{>{\centering\arraybackslash}p{#1}}

\usepackage{amssymb}
\usepackage{graphicx}
\usepackage{grffile}
\usepackage{youngtab}
\Yboxdim{6pt}
\input epsf

\makeatletter\@addtoreset{equation}{section}\makeatother

\def\bZ {\mathbb{Z}}

\def\be{\begin{equation}}
\def\ee{\end{equation}}
\def\bea{\begin{eqnarray}}
\def\eea{\end{eqnarray}}
\def\ie{\begin{equation}\begin{aligned}}
\def\fe{\end{aligned}\end{equation}}

\newcommand{\m}{\mu}

\newcommand{\A}{{\alpha}}

\makeatletter\@addtoreset{equation}{section}\makeatother

\newcommand{\vev}[1]{{\left< {#1} \right>}}

\newcommand{\cD}{{\mathcal D}}

\newcommand{\cN}{{\mathcal N}}
\newcommand{\cO}{{\mathcal O}}

\renewcommand{\title}[1]{\vbox{\center\LARGE{#1}}\vspace{5mm}}
\renewcommand{\author}[1]{\vbox{\center#1}\vspace{5mm}}
\newcommand{\address}[1]{\vbox{\center\em#1}}
\newcommand{\email}[1]{\vbox{\center\tt#1}\vspace{5mm}}

\begin{document}

\unitlength = .8mm
\begin{titlepage}
\begin{center}
\hfill \\
\hfill \\
\vskip 1cm

\title{A semi-local holographic minimal model}

\author{Chi-Ming Chang$^{a}$ and Xi Yin$^{b}$}

\address{Jefferson Physical Laboratory, Harvard University,\\
Cambridge, MA 02138 USA}

\email{$^a$cmchang@physics.harvard.edu,
$^b$xiyin@fas.harvard.edu}

\end{center}

\abstract{We present a conjecture on the complete spectrum of single-trace operators in the infinite $N$ limit of $W_N$ minimal model and evidences for the conjecture. We further propose that the holographic dual of $W_N$ minimal model in the 't Hooft limit is an unusual ``semi-local" higher spin gauge theory on $AdS_3\times S^1$. At each point on the $S^1$ lives a copy of three-dimensional Vasiliev theory, that contains an infinite tower of higher spin gauge fields coupled to a single massive complex scalar propagating in $AdS_3$. The Vasiliev theories at different points on the $S^1$ are correlated only through the $AdS_3$ boundary conditions on the massive scalars. All but one single tower of higher spin symmetries are broken by the boundary conditions.}

\vfill

\end{titlepage}

\eject 

\section{Introduction}

The AdS/CFT correspondence \cite{Maldacena:1997re} in principle gives a precise and non-perturbative formulation of quantum gravity in terms of large $N$ gauge theories. In practice, our understanding of quantum gravity using AdS/CFT has been largely limited by difficulties in solving strongly coupled large $N$ gauge theories. Thus, exactly solvable models of strongly coupled gauge theories with a semi-classical gravity dual are highly desirable. In two dimensions, there are lots of exactly solvable conformal field theories. Most of them do not have a large $N$ limit that allows for a weakly coupled gravity dual. In \cite{Gaberdiel:2010pz}, Gaberdiel and Gopakumar proposed that $W_N$ minimal model, which may be constructed as the coset model
$$
{SU(N)_k\times SU(N)_1\over SU(N)_{k+1}}
$$
has a 't Hooft-like large $N$ limit, where $N,k$ are taken to infinity, while the 't Hooft coupling
$$
\lambda = {N\over k+N}
$$
is held fixed (between 0 and 1). The $W_N$ minimal model was conjectured to be dual to a higher spin gauge theory in $AdS_3$, coupled to massive scalar fields. It was observed that, in particular, a class of $W_N$ primary operators in the CFT has order 1 conformal weight in the large $N$ limit.
The central charge of the CFT is
\ie
c&=(N-1)\left(1-{N(N+1)\over (N+k)(N+k+1)}\right) =N(1-\lambda^2)+\cO(N^0).
\fe
The linear dependence on $N$ is characteristic of a vector model. It was not at all obvious, however, that the primary operators can be classified as single-trace or multi-trace operators in the large $N$ limit, which are supposed to be dual to single elementary particle states or their bound states in the bulk. It was also not obvious that the correlation functions of the primaries obey large $N$ factorization, according to the appropriate classification of single-trace and multi-trace operators. There is the further complication with the existence of a large set of ``light primaries", whose conformal dimension go to zero in the infinite $N$ limit, which nonetheless do not decouple in the correlation functions \cite{Papadodimas:2011pf}.

The identification of a number of single-trace versus multi-trace operators, and the large $N$ factorization of their correlation functions, are demonstrated in \cite{Chang:2011vka} by analyzing exact results of three-point functions. There, a complete picture of the single-trace/one-particle spectrum and the interpretation of the single-trace light primaries was still missing. In this paper, we will present a conjecture on the complete spectrum of single-trace operators of finite conformal weight in the infinite $N$ limit. Importantly, we will argue that there are a large set of hidden symmetries in $W_N$ minimal model that emerge in the infinite $N$ limit, and are broken by $1/N$ effects, in a manner that is similar to higher spin symmetries in three-dimensional Chern-Simons vector models. The currents that generate these hidden symmetries come from $W_N$ descendants of light primaries at finite $N$, but effectively become primary fields in the infinite $N$ limit. They are dual to gauge fields in $AdS_3$ of various spins, under which the massive scalars are charged. The corresponding gauge transformations are incompatible with the boundary conditions on the massive scalars, which leads to the breaking of symmetry.

Our conjecture on the large $N$ spectrum, combined with the identification of the gauge generators acting on the matter scalars, leads to a dramatically new picture of the holographic dual of the $W_N$ minimal model. We propose that the dual higher spin gauge theories is a ``semi-local"\footnote{The terminology comes from analogy with the holographic theory of semi-local quantum liquids \cite{Iqbal:2011in}.} theory living on $AdS_3\times S^1$. This is not an ordinary four-dimensional field theory, however. At each point of the $S^1$, there is a tower of higher spin gauge fields in $AdS_3$, coupled to a single complex massive scalar field, of the type described by Vasiliev's system in three dimensions. The different Vasiliev theories at different points on the $S^1$ appear to be decoupled at the level of bulk equations of motion. Rather, they interact only through the boundary condition which mixes scalar fields living at different points on the circle $S^1$. Essentially, while all the scalars classically have the same mass in $AdS_3$, the boundary condition assigns one scaling dimension $\Delta_+$ on right-moving modes of the scalar on the circle, and the complementary scaling dimension $\Delta_-=2-\Delta_+$ on left-moving modes of the scalar on the circle.

While our proposal for the holographic dual is rather unconventional due to the large degeneracy in the bulk fields, it seems to be unavoidable due to peculiarities in the structure of large $N$ factorization in $W_N$ minimal model. We believe that it is characteristic of gauged vector models on non-simply connected spaces \cite{Banerjee:2012gh,Banerjee:2012aj}. Presumably, what we see here is the field theory of the tensionless limit of a more conventional string theory in $AdS_3$, dual to quiver-like generations of the $W_N$ minimal model, and the $S^1$ should come from a topological sector of the string theory in this limit.

In the next section, we briefly review the construction of $W_N$ minimal model CFT, and what's previously known about the structure of single-trace and multi-trace operators in this CFT in the infinite $N$ limit. In section 3 and 4, we present some new examples of single-trace operators and operator relations involving light primaries at large $N$. In section 5, we argue that the operator relations that seemed to be in conflict with large $N$ factorization should in fact be interpreted as current non-conservation relations for currents that generate approximate ``hidden" symmetries in the large $N$ limit. Further data on higher spin currents of this sort are presented in section 6. In section 7, we state our conjecture on the complete spectrum of single-trace operators in the CFT at infinite $N$, or single-particle states in the bulk. These include the infinite family of massive scalars $\phi_n, \tilde\phi_n$, light scalars $\omega_n$, and the hidden higher spin currents $j_n^{(s)}$, all of which are complex. Various checks based on partition functions and characters are given by section 8. In section 9, we determine the gauge generators associated with the hidden symmetry currents, and reveal the picture of semi-local higher spin theory on $AdS_3\times S^1$. We discuss the implication of our results in section 10.

\section{Summary of previous results}

The $W_N$ minimal model has a holomorphic higher spin current $W^{(s)}$ and an anti-holomorphic current $\overline W^{(s)}$ for each spin $s=2,3,4,\cdots,N$. The Fourier modes of $W^{(s)}$ generate the $W_N$ algebra, which is a higher spin generalization of the Virasoro algebra. In the large $N$ limit, the $W_N$ algebra turns into the $W_\infty[\lambda]$ algebra that contains generators with arbitrary spins. In $W_N$ minimal model, the $W_N$ primary operators, the primaries with respect to the $W_N$ algebra, can be labeled by two representations $(\Lambda_+,\Lambda_-)$, where $\Lambda_\pm$ are the highest weight representations of $SU(N)_k$ and $SU(N)_{k+1}$, respectively.\footnote{A prior, the primary should also depend on the highest weight representation $\Lambda_0$ of $SU(N)_1$. However, $\Lambda_0$ can be determining by requiring $\Lambda_++\Lambda_0-\Lambda_-$ being inside the root lattice of $SU(N)$.} For fixed representations $\Lambda_+, \Lambda_-$ at sufficiently large $N$,\footnote{Namely representations that are found in the tensor product of finitely many fundamental or anti-fundamental representations of $SU(N)$, at large $N$.} the fusion coefficients for the primary operators in the $W_N$ minimal model is simply given by the product of the fusion coefficients in the $SU_k$ and $SU_{k+1}$ WZW models, i.e.
\ie\label{FRC}
\cN^{W_N}_{(\Lambda_+^1,\Lambda_-^1)(\Lambda_+^2,\Lambda_-^2)}{}^{(\Lambda_+^3,\Lambda_-^3)}=\cN^{(k)}_{\Lambda_+^1\Lambda^2_+}{}^{\Lambda_+^3}\cN^{(k+1)}_{\Lambda_-^1\Lambda^2_-}{}^{\Lambda_-^3},
\fe
where $\cN^{(k)}_{\Lambda^1\Lambda^2}{}^{\Lambda^3}$ is the fusion coefficient of $SU(N)_k$ WZW model.

The gravity dual of $W_N$ minimal model at large $N$ must be a higher spin gauge theory, containing a tower of gauge fields of spins $s=2,3,4,\cdots, \infty$ that are dual to the higher spin currents $W^{(s)}$. The pure higher spin gauge theory on $AdS_3$ can be described by the Chern-Simons action with $hs(\lambda)\times hs(\lambda)$ gauge algebra. The higher spin algebra $hs(\lambda)$ is an infinite dimensional Lie algebra, and by a Brown-Henneaux type computation, it was shown, in \cite{Henneaux:2010xg,Campoleoni:2010zq,Gaberdiel:2011wb}, that the asymptotic symmetry algebra of $hs(\lambda)$ is $W_\infty[\lambda]$. It also follows from this computation that the bulk coupling constant is proportional to inverse the square root of the central charge, i.e.
\ie
g_{bulk}\sim{1\over \sqrt{c}}\sim{1\over \sqrt{N}}.
\fe
The primary operators in the $W_N$ minimal model, constructed from the diagonal modular invariant, do not carry spin. They should be dual to scalar elementary particles and their bound states with zero angular momentum, that become unbound in the infinite $N$ (zero bulk coupling) limit. In particular, the primary operator $\phi_1=(\yng(1),0)$ is dual to a scalar field with left and right conformal weight equal to
\ie
h_{(\yng(1),0)}={1\over 2}(1+\lambda)
\fe
in the large $N$ limit. The primary $\bar\phi_1=(\bar{\yng(1)},0)$ has the same dimension in the large $N$ limit, and is dual to the anti-particle of $(\yng(1),0)$. The primary operators $(\yng(1,1),0)$ and $(\yng(2),0)$ have conformal weights
\ie
h_{(\yng(1,1),0)}=1+\lambda,~~~h_{(\yng(2),0)}=2+\lambda
\fe
in the large $N$ limit. Note that $h_{(\yng(1,1),0)}$ and $h_{(\yng(2),0)}$ are twice the dimension of $(\yng(1),0)$ plus a non-negative integer. This allows for the identification of $(\yng(1,1),0)$ and $(\yng(2),0)$ as two-particle states of $\phi_1$'s. In general, the primary operators of the form $(\Lambda,0)$ are dual to the multi-particle states of $B(\Lambda)$ $\phi_1$'s, where $B(\Lambda)$ is the number of boxes of the Young tableaux of the representation $\Lambda$ (here we assume that $B(\Lambda)$ does not scale with $N$). The $W_N$ minimal model in the large $N$ limit has a symmetry that exchanges $\Lambda_+$ with $\Lambda_-$, while flipping the sign of $\lambda$. Hence, the primary $\tilde\phi_1=(0,\yng(1))$ is also dual to a scalar elementary particle, with dimension
\ie
h_{(0,\yng(1))}={1\over 2}(1-\lambda),
\fe
and the primaries $(0,\Lambda)$ are dual to the multi-particle states of $\tilde\phi_1$. The fusion coefficients \eqref{FRC} implies that the primaries of the form $(\Lambda,0)$ (or $(0,\Lambda)$) are closed under the OPE, as long as $\Lambda$ is small compared to $N$. They form a closed subsector of the $W_N$ minimal model in the large $N$ limit. Either one of these two subsectors has a consistent set of $n$-point functions on the sphere, in the sense that they factorize through only operators within the same subsector. In \cite{Chang:2011mz}, we proposed a bulk dual for each of the subsectors. The classical bulk theory is described by Vasiliev's system in three dimensions \cite{Prokushkin:1998bq,Prokushkin:1998vn,Chang:2011mz}, which is a higher spin gauge theory of gauge fields of spin $s = 2, 3,\cdots,\infty$ based on the higher spin algebra $hs(\lambda)$, coupled to a complex massive scalar field of mass squared $m^2=-(1-\lambda^2)$. This conjecture has also been checked by computations of the three-point function $\vev{\phi_1\bar\phi_1W^{(s)}}$ on both side of the correspondence \cite{Chang:2011mz,Ammon:2011ua}. The bulk dual of other primary operators was first studied in \cite{Papadodimas:2011pf}, and later extended in \cite{Chang:2011vka}.

In \cite{Chang:2011vka}, we computed the three-point functions of $W_N$ primaries $(\yng(1),0)$, $(\yng(1),\yng(1))$, and/or their charge conjugates, with the primary $(\Lambda_+,\Lambda_-)$ where $\Lambda_\pm$ are $\yng(2)$ or $\yng(1,1)$. This result allowed us to identify the primary operators $(\Lambda_+,\Lambda_-)$, for $\Lambda_\pm$ being one- or two-box representations, with the single-particles or multi-particle states in the bulk in large $N$ limit. The result in \cite{Chang:2011vka} can be summarized in the following table:

\noindent\begin{tabular}{|x{1.2cm}|x{0.4cm}|x{2.4cm}|x{4.6cm}|x{4.6cm}|}\hline
\diag{0cm}{1.2cm}{~~~$\Lambda_+$}{$\Lambda_-$} & 0 & $\yng(1)$ & $\yng(2)$ & $\yng(1,1)$ 
\\ \hline
0 & 0 & $\tilde\phi_1$ & $L_{\tilde\phi_1}$ & $\tilde\phi_1^2$
\\ \hline
$\yng(1)$ & $\phi_1$ & $\omega_1$ & ${1\over \sqrt{2}}(\tilde\phi_1\omega_1+\tilde\phi_2)$ & ${1\over \sqrt{2}}(\tilde\phi_1\omega_1-\tilde\phi_2)$  
\\ \hline
$\yng(2)$ & $L_{\phi_1}$ & ${1\over \sqrt{2}}(\phi_1\omega_1+\phi_2)$ & ${1\over 2}(\omega^2_1+\sqrt{2}\omega_2)$ & ${1\over \sqrt{2}}(L_{\omega_1}-{1\over \sqrt{2}}(\phi_1\tilde\phi_2-\phi_2\tilde\phi_1))$ 
\\ \hline
$\yng(1,1)$ & $\phi_1^2$ & ${1\over \sqrt{2}}(\phi_1\omega_1-\phi_2)$ & ${1\over \sqrt{2}}(L_{\omega_1}+{1\over \sqrt{2}}(\phi_1\tilde\phi_2-\phi_2\tilde\phi_1))$  & ${1\over 2}(\omega^2_1-\sqrt{2}\omega_2)$ 
\\ \hline
\end{tabular}

\noindent where the $\phi_1,\tilde\phi_1,\omega_1,\phi_2,\tilde\phi_2,\omega_2$ are operators that dual to the elementary particles in the bulk:
\ie
&\phi_1=(\yng(1),0),~~~\tilde\phi_1=(0,\yng(1)),~~~\omega_1=(\yng(1),\yng(1)),
\\
&\phi_2={1\over \sqrt{2}}\left[(\yng(2),\yng(1))-(\yng(1,1),\yng(1))\right],~~~\tilde\phi_2={1\over \sqrt{2}}\left[(\yng(1),\yng(2))-(\yng(1),\yng(1,1))\right],
\\
&\omega_2={1\over \sqrt{2}}\left[(\yng(2),\yng(2))-(\yng(1,1),\yng(1,1))\right].
\fe
Two comments about this identification: first note that the expressions only make sense in the large $N$ limit since each term in the linear combination has different dimension in the subleading order of $1/N$. In the large $N$ limit, we conjecture that each term in the above linear combination has the same dimensions and higher spin charges. This conjecture has been checked up to spin 5; see appendix A.  Second, in the table, the products of the operators are well-defined because one can check that the OPE's of the them have no singularity in the large $N$ limit. The operator $L_{\cO}$ is defined as
\ie
&L_{\cO}={1\over 2 \sqrt{2}h_{\cO}}\left(\cO\partial\bar\partial\cO-\partial \cO \bar\partial\cO\right).
\fe
Again, the products are well-defined since there is no singularity in the OPE. This table is further subject to a relation \cite{Papadodimas:2011pf}:
\ie\label{OMR1}
&{1\over 2h_{\omega_1}}\partial\bar\partial\omega_1= \phi_1\tilde\phi_1.
\fe
The bulk physical meaning of this relation will be explain in detail in the section 5.

\section{New single-trace operators/elementary particles}

Let us extend this table to the the representation with three boxes. Before diving into the computation of three-point functions, there are some principles can help us to determine whether a primary operator $\cO_A$ can be dual to the two-particle state of two elementary particles that are dual to $\cO_B$ and $\cO_C$. First, the primary $\cO_A$ must appear in the OPE of the primary $\cO_B$ and $\cO_C$. Second, the dimension of the primary $\cO_A$ must be equal to the sum of the dimension of $\cO_B$ and $\cO_C$ up to higher order corrections in $1/N$. Following is a table summarizing the dimension of the primary operator up to representation of three boxes. 

\noindent\begin{tabular}{|x{1.2cm}|x{1.9cm}|x{1.7cm}|x{1.7cm}|x{1.2cm}|x{1.9cm}|x{1.9cm}|x{1.2cm}|}\hline
\diag{0cm}{1.2cm}{~~~$\Lambda_+$}{$\Lambda_-$} & 0 & $\yng(1)$ & $\yng(2)$ & $\yng(1,1)$ & $\yng(3)$ & $\yng(2,1)$ & $\yng(1,1,1)$ 
\\ \hline
0 & 0 & $1-\lambda\over 2$ & $(1-\lambda)+1$ & $1-\lambda$ & $3\left({1-\lambda\over 2}\right)+3$ & $3\left({1-\lambda\over 2}\right)+1$ & $3\left({1-\lambda\over 2}\right)$
\\ \hline
$\yng(1)$ & $1+\lambda\over 2$ & ${\lambda^2\over 2N}$ & ${1-\lambda\over 2}$ & ${1-\lambda\over 2}$  & $(1-\lambda)+1$ & $1-\lambda$ & $1-\lambda$ 
\\ \hline
$\yng(2)$ & $(1+\lambda)+1$ & ${1+\lambda\over 2}$ & ${\lambda^2\over N}$ & 1 & ${1-\lambda\over 2}$ & ${1-\lambda\over 2}$ & ${1-\lambda\over 2}+1$ 
\\ \hline
$\yng(1,1)$ & $1+\lambda$ & ${1+\lambda\over 2}$ & 1 & ${\lambda^2\over N}$ & ${1-\lambda\over 2}+2$ & ${1-\lambda\over 2}$ & ${1-\lambda\over 2}$ 
\\ \hline
$\yng(3)$ & $3\left({1+\lambda\over 2}\right)+3$ & $(1+\lambda)+1$ & ${1+\lambda\over 2}$ & ${1+\lambda\over 2}+2$ & ${3\lambda^2\over 2N}$ & 1 & 3 
\\ \hline
$\yng(2,1)$ & $3\left({1+\lambda\over 2}\right)+1$ & $1+\lambda$ & ${1+\lambda\over 2}$ & ${1+\lambda\over 2}$ & 1 & ${3\lambda^2\over 2N}$ & 1 
\\ \hline
$\yng(1,1,1)$ & $3\left({1+\lambda\over 2}\right)$ & $1+\lambda$ & ${1+\lambda\over 2}+1$ & ${1+\lambda\over 2}$ & 3 & 1 & ${3\lambda^2\over 2N}$
\\ \hline
\end{tabular}

Let us first focus on the light states: $(\yng(3),\yng(3)),(\yng(2,1),\yng(2,1)),(\yng(1,1,1),\yng(1,1,1))$. By the fusion rule and the additivity of the dimension, two linear combinations of these three operators can be identified with the multi-particle states $\omega_1^3$ and $\omega_1\omega_2$. Let us see this explicitly in terms of structure constants. A formula of a large class of the structure constants is given in \cite{Chang:2011vka}. By explicitly evaluating the formula, we find out that, in the large $N$ limit, the OPE of $(\yng(1),\yng(1))$ and $(\yng(2),\yng(2))$ has no singularity, hence the product $(\yng(1),\yng(1))(\yng(2),\yng(2))$ is well-defined, which in the large $N$ limit is
\ie
&(\yng(1),\yng(1))(\yng(2),\yng(2))=(\yng(3),\yng(3))+(\yng(2,1),\yng(2,1)).
\fe
Similarly, in the large $N$ limit, we have
\ie
&(\yng(1),\yng(1))(\yng(1,1),\yng(1,1))=(\yng(2,1),\yng(2,1))+(\yng(1,1,1),\yng(1,1,1)).
\fe
Rewriting the equation in terms of $\omega_1,\omega_2$, we have
\ie
\omega_1\omega_2&=(\yng(3),\yng(3))-(\yng(1,1,1),\yng(1,1,1)),
\\
\omega_1^3&=(\yng(3),\yng(3))+2(\yng(2,1),\yng(2,1))+(\yng(1,1,1).\yng(1,1,1)).
\fe
There is one linear combination of $(\yng(3),\yng(3)),(\yng(2,1),\yng(2,1)),(\yng(1,1,1),\yng(1,1,1))$, which cannot be expressed as $\omega_1\omega_2,\omega_1^3$. This operator should be dual to a new light elementary particle. Hence, we define
\ie
\omega_3&={1\over \sqrt{3}}\left[(\yng(3),\yng(3))-(\yng(2,1),\yng(2,1))+(\yng(1,1,1),\yng(1,1,1))\right],
\fe
which is orthonormal to $\omega_1\omega_2,\omega_1^3$ and is a new elementary light particle.

Next, let us look at the primaries with dimension ${1-\lambda\over 2}$ and with three boxes representations. They are $(\yng(2),\yng(3)),(\yng(2),\yng(2,1)),(\yng(1,1),\yng(2,1)),(\yng(1,1),\yng(1,1,1)).$ From the additivity of the dimension, three linear combinations of these four operators can be dual to the multi-particle states $\tilde\phi_1\omega_2,\tilde\phi_1\omega_1^2,\tilde\phi_2\omega_1$. Again, we can see this explicitly from the structure constants. From the structure constant computation, we have the following products at large $N$:
\ie
&(0,\yng(1))(\yng(2),\yng(2))={1\over \sqrt{3}}(\yng(2),\yng(3))+\sqrt{2\over 3}(\yng(2),\yng(2,1)),
\\
&(0,\yng(1)) (\yng(1,1),\yng(1,1))=\sqrt{2\over 3}(\yng(1,1),\yng(2,1))+{1\over \sqrt{3}}(\yng(1,1),\yng(1,1,1)),
\\
&(\yng(1),\yng(1))(\yng(1),\yng(2))=\sqrt{2\over 3}(\yng(2),\yng(3))+{1\over 2\sqrt{3}}(\yng(2),\yng(2,1))+{\sqrt{3}\over 2}(\yng(1,1),\yng(2,1)),
\\
&(\yng(1),\yng(1))(\yng(1),\yng(1,1))={\sqrt{3}\over 2}(\yng(2),\yng(2,1))+{1\over 2\sqrt{3}}(\yng(1,1),\yng(2,1))+\sqrt{2\over 3}(\yng(1,1),\yng(1,1,1)).
\fe
Expressing them in terms of $\tilde\phi_1,\tilde\phi_2,\omega_1,\omega_2$, we obtain
\ie
\tilde\phi_1\omega_2&={1\over\sqrt{6}}\left[(\yng(2),\yng(3))+\sqrt{2}(\yng(2),\yng(2,1))-\sqrt{2}(\yng(1,1),\yng(2,1))-(\yng(1,1),\yng(1,1,1))\right],
\\
{1\over \sqrt{2}}\tilde\phi_1\omega^2_1&=\tilde\phi_1{(\yng(2),\yng(2))+(\yng(1,1),\yng(1,1))\over \sqrt{2}}={1\over\sqrt{6}}\left[(\yng(2),\yng(3))+\sqrt{2}(\yng(2),\yng(2,1))+\sqrt{2}(\yng(1,1),\yng(2,1))+(\yng(1,1),\yng(1,1,1))\right]
\\
&={1\over \sqrt{2}}\omega_1{\left[(\yng(1),\yng(2))+(\yng(1),\yng(1,1))\right]\over \sqrt{2}}={1\over\sqrt{6}}\left[(\yng(2),\yng(3))+\sqrt{2}(\yng(2),\yng(2,1))+\sqrt{2}(\yng(1,1),\yng(2,1))+(\yng(1,1),\yng(1,1,1))\right],
\\
\tilde\phi_2\omega_1&={1\over\sqrt{6}}\left[\sqrt{2}(\yng(2),\yng(3))-(\yng(2),\yng(2,1))+(\yng(1,1),\yng(2,1))-\sqrt{2}(\yng(1,1),\yng(1,1,1))\right].
\fe
There is one linear combination of $(\yng(2),\yng(3)),(\yng(2),\yng(2,1)),(\yng(1,1),\yng(2,1)),(\yng(1,1),\yng(1,1,1))$, which is linear independent of $\tilde\phi_1\omega_2,\tilde\phi_1\omega_1^2,\tilde\phi_2\omega_1$, and should be dual to a new elementary particle in the bulk. Hence, we can define
\ie
\tilde\phi_3={1\over \sqrt{6}}\left[\sqrt{2}(\yng(2),\yng(3))-(\yng(2),\yng(2,1))-(\yng(1,1),\yng(2,1))+\sqrt{2}(\yng(1,1),\yng(1,1,1))\right],
\fe
which is orthonormal to $\tilde\phi_1\omega_2,{1\over \sqrt{2}}\tilde\phi_1\omega^2_1,\tilde\phi_2\omega_1$. Similarly, by exchanging the left and right representations, we have
\ie
\phi_1\omega_2&={1\over\sqrt{6}}\left[(\yng(3),\yng(2))+\sqrt{2}(\yng(2,1),\yng(2))-\sqrt{2}(\yng(2,1),\yng(1,1))-(\yng(1,1,1),\yng(1,1))\right],
\\
{1\over \sqrt{2}}\phi_1\omega^2_1&={1\over\sqrt{6}}\left[(\yng(3),\yng(2))+\sqrt{2}(\yng(2,1),\yng(2))+\sqrt{2}(\yng(2,1),\yng(1,1))+(\yng(1,1,1),\yng(1,1))\right],
\\
\phi_2\omega_1&={1\over\sqrt{6}}\left[\sqrt{2}(\yng(3),\yng(2))-(\yng(2,1),\yng(2))+(\yng(2,1),\yng(1,1))-\sqrt{2}(\yng(1,1,1),\yng(1,1))\right],
\fe
and we define
\ie
\phi_3={1\over \sqrt{6}}\left[\sqrt{2}(\yng(3),\yng(2))-(\yng(2,1),\yng(2))-(\yng(2,1),\yng(1,1))+\sqrt{2}(\yng(1,1,1),\yng(1,1))\right].
\fe

Next, let us focus on the primaries $(\yng(1),\yng(2,1)),(\yng(1),\yng(1,1,1))$. By the fusion rule and the additivity of the dimension, it is not hard to see that they must be identified with the two linear combinations of $\tilde\phi_1\tilde\phi_2$ and $\omega_1\tilde\phi_1^2$, which are dual to two- and three-particle states. Similarly, the primaries $(\yng(2,1),\yng(1)),(\yng(1,1,1),\yng(1))$ are identified with the two linear combinations of $\phi_1\phi_2$ and $\omega_1\phi_1^2$. All the other primaries: $(\yng(1),\yng(3))$, $(\yng(2),\yng(1,1,1))$, $(\yng(1,1),\yng(3))$, $(\yng(3),\yng(2,1))$, $(\yng(3),\yng(1,1,1))$, $(\yng(2,1), \yng(1,1,1))$, and the primaries with left and right representations exchanged, are also dual to multi-particle states. We will show this in section 8.

\section{Large $N$ operator relations involving $\omega_2$ and $\omega_3$}

There is a new relation involving the descendant of $\omega_2$, similar to the relation \eqref{OMR1}. By the following two structure constants:
\ie
&C_{nor}\left((0,\yng(1)),(\yng(1,1),\yng(1)),(\overline{\yng(1,1)},\overline{\yng(1,1)})\right)={\sqrt{2}\over N}+\cO({1\over N^2}),
\\
&C_{nor}\Big((0,\yng(1)),(\yng(2),\yng(1)),(\overline{\yng(2)},\overline{\yng(2)})\Big)={\sqrt{2}\over N}+\cO({1\over N^2}),
\fe
we have the three-point functions:
\ie\label{OPhPh2}
\vev{\bar\omega_2(z)\phi_1(w)\tilde\phi_2(0)}=\vev{\bar\omega_2(z)\phi_2(w)\tilde\phi_1(0)}={1\over \sqrt{2}N}{1\over |z-w|^{2\lambda}|w|^2|z|^{-2\lambda}},
\fe
in the large $N$ limit. Taking $\partial\bar\partial$ on $\bar\omega_2$, we obtain:
\ie\label{OOPPLL}
\vev{\partial\bar\partial\bar\omega_2(z)\phi_1(w)\tilde\phi_2(0)}=\vev{\partial\bar\partial\bar\omega_2(z)\phi_2(w)\tilde\phi_1(0)}={\lambda^2\over \sqrt{2}N}\left({1\over |z-w|^{2(1+\lambda)}}\right)\left({1\over |z|^{2(1-\lambda)}}\right).
\fe
The two factors on the right hand side of \eqref{OOPPLL} are precisely given by the two-point functions of $\vev{\phi_2\bar\phi_2}$ and $\vev{\tilde\phi_1\bar{\tilde\phi}_1}$, or $\vev{\phi_1\bar\phi_1}$ and $\vev{\tilde\phi_2\bar{\tilde\phi}_2}$. Hence, this suggests the following relation in the large $N$ limit:
\ie\label{OMR2}
{1\over 2h_{\omega_2}}\partial\bar\partial\omega_2={1\over \sqrt{2}}(\phi_1\tilde\phi_2+\tilde\phi_1\phi_2).
\fe
To make sure that there are no extra term on the left hand side, one can compute the two-point function for the right hand side of \eqref{OMR2} with its charge conjugate, and the two-point function for the left hand side of \eqref{OMR2} with its charge conjugate, and find that they agree.

Form the previous analysis on $\omega_1,\omega_2$, it suggests that there is also a relation involving the descendant of $\omega_3$. We postulate such relation should be
\ie\label{OMR3}
{1\over 2h_{\omega_3}}\partial\bar\partial \omega_3= {1\over \sqrt{3}}(\phi_1\tilde\phi_3+\phi_2\tilde\phi_2+\phi_3\tilde\phi_1).
\fe
We give an argument for this relation. In the large $N$ limit, we have the following structure constants
\ie
&C_{nor}\Big((0,\yng(1)),(\yng(3),\yng(2)),(\overline{\yng(3)},\overline{\yng(3)})\Big)={\sqrt{3}\over N},~~~C_{nor}\Big((0,\yng(1)),(\yng(1,1,1),\yng(1,1)),(\overline{\yng(1,1,1)},\overline{\yng(1,1,1)})\Big)={\sqrt{3}\over N}
\\
&C_{nor}\Big((0,\yng(1)),(\yng(2,1),\yng(2)),(\overline{\yng(2,1)},\overline{\yng(2,1)})\Big)=\sqrt{3\over 2}{1\over N},~~~C_{nor}\Big((0,\yng(1)),(\yng(2,1),\yng(1,1)),(\overline{\yng(2,1)},\overline{\yng(2,1)})\Big)=\sqrt{3\over 2}{1\over N}.
\fe
These structure constants give the three-point functions:
\ie\label{OPhPh3}
\vev{\bar\omega_3(z)\phi_1(w)\tilde\phi_3(0)}=\vev{\bar\omega_3(z)\phi_3(w)\tilde\phi_1(0)}={1\over \sqrt{3}N}{1\over |z-w|^{2\lambda}|w|^2|z|^{-2\lambda}},
\fe
in the large $N$ limit. Taking $\partial\bar\partial$ on $\bar\omega_2$, the three-point function again factorizes as a product of two two-point functions:
\ie\label{TPFFF}
\vev{\partial\bar\partial\bar\omega_3(z)\phi_1(w)\tilde\phi_3(0)}=\vev{\partial\bar\partial\bar\omega_3(z)\phi_3(w)\tilde\phi_1(0)}={\lambda^2\over \sqrt{3}N}{1\over |z-w|^{2(1+\lambda)}|z|^{2(1-\lambda)}}.
\fe
The three-point functions \eqref{TPFFF} imply the relation
\ie
{1\over 2h_{\omega_3}}\partial\bar\partial \omega_3= {1\over \sqrt{3}}(\phi_1\tilde\phi_3+\phi_3\tilde\phi_1+\cdots).
\fe
By comparing the two-point functions of the left and right hand sides with their charge conjugates, we know that the ``$\cdots$" must take the form as a single term $\phi_n\tilde\phi_m$ with $n,m\neq 1,3$, and the only candidate is $\phi_2\tilde\phi_2$.

\section{Hidden symmetries}

In this section, we give physical interpretation of the relations \eqref{OMR1}, \eqref{OMR2}, \eqref{OMR3}, and provide a bulk mechanism of producing such relations. The key observation is that the dimension of $\omega_n$ goes to zero in the large $N$ limit. Therefore, it should effectively behave like a free boson, whose derivative is a conversed current. Hence, we define the holomorphic current $(j^{(1)}_n)_z=\partial\omega_n/\sqrt{2h_{\omega_n}}$ and also the antiholomorphic current $(j^{(1)}_n)_{\bar z}=\bar\partial\omega_n/\sqrt{2h_{\omega_n}}$, for $n=1,2,3$, which has normalized two-point function with itself. For simplicity, we will sometimes suppress the index by simply denoting $(j^{(1)}_n)_{ z}$ as $j^{(1)}_n$ in the following. However, since the dimensions of $\omega_n$ are not exactly equal to zero, the currents $j_n^{(1)}$ are not exactly conserved. The relations \eqref{OMR1}, \eqref{OMR2}, \eqref{OMR3} are then naturally interpreted as current non-conservation equations\footnote{The current non-conservations equation for theories in one higher dimension have been studied in \cite{Giombi:2011kc,Maldacena:2012sf,Chang:2012kt}.}:
\ie\label{CNEn}
\bar\partial j_n^{(1)}={\lambda\over\sqrt{N}}(\phi_1\tilde \phi_n+\phi_2\tilde \phi_{n-1}+\cdots+\phi_n\tilde\phi_1).
\fe

The bulk interpretation of these current non-conservation equations is simple. Let us illustrate this by considering the case of $j_n^{(1)}$. In this case the current non-conservation equation is simply
\ie\label{CNE1}
\bar\partial j_1^{(1)}={\lambda\over\sqrt{N}}\phi_1\tilde \phi_1.
\fe
Following the $AdS/CFT$ dictionary, the bulk dual of the current $j_1^{(1)}$ is a $U(1)$ Chern-Simons gauge field $A_\m$, and the bulk dual of the operators $\phi_1,\tilde\phi_1$ are two scalars $\Phi,\widetilde\Phi$. These two scalars have different but complementary dimensions, hence they have the same mass but different boundary conditions. They can be minimally coupled to the gauge field $A_\m$. The action of this system up to cubic order is
\ie
S&={k_{CS}\over 4\pi}\int  AdA+2i\int d^2xdz\sqrt{g}  A^\m\left[\widetilde\Phi \partial_\m \Phi-\Phi \partial_\m \widetilde\Phi\right].
\fe
Using this action, we can compute the three-point function of $\bar\partial j_1^{(1)}$ with $\phi_1,\tilde\phi_1$. The boundary to bulk propagator of the Chern-Simons gauge field takes a pure gauge form $A_\m=\partial_\m\Lambda$. The cubic action, hence, can be written as
\ie
\lim_{z\to 0} {2\over z}\int d^2x  \Lambda\left[\Phi \partial_z \widetilde\Phi-\widetilde\Phi \partial_z \Phi\right].
\fe
The three-point function is then given by
\ie
&\vev{ j_1^{(1)}(\vec x_3)\phi_1(x_1)\tilde\phi_1(x_2)}
\\
&=\lim_{z\to 0} {2\over z}\int d^2x\Lambda(x-x_3)\big[K_{1+\lambda}(x-x_1) \partial_z K_{1-\lambda}(x-x_2)-K_{1-\lambda}(x-x_2) \partial_z K_{1+\lambda}(x-x_1)\big]
\\
&= -16\pi \lambda\int d^2x  {1\over (x^+-x_3^+)} {1\over |\vec x-\vec x_2|^{2(1-\lambda)}} {1\over |\vec x-\vec x_1|^{2(1+\lambda)}},
\fe
where $K_\Delta$ and $\Lambda$ are the boundary to bulk propagators for the scalar and gauge function:
\ie
K_\Delta=\left(z\over z^2+|\vec x|^2\right)^\Delta,~~~\Lambda={4\pi\over x^+}.
\fe
Taking the derivative ${\partial\over \partial x^+_3}$ on the above expression, we obtain
\ie\label{cstpf}
\vev{\bar\partial j_1^{(1)}(\vec x_3)\phi_1(x_1)\tilde\phi_1(x_2)}&=-16\pi^2 \lambda\int d^2x  \delta^2(x-x_3) {1\over |\vec x-\vec x_2|^{2(1-\lambda)}} {1\over |\vec x-\vec x_1|^{2(1+\lambda)}}
\\
&=-16\pi^2 \lambda{1\over |\vec x_3-\vec x_2|^{2(1-\lambda)} |\vec x_3-\vec x_1|^{2(1+\lambda)}},
\fe
which factories into a product of two two-point functions of scalars with dimension $\Delta=1+\lambda$ and $1-\lambda$. This matches exactly with what we expected from \eqref{CNE1} provided the identification of the Chern-Simons level $k_{CS}=N$. In section 8, we will show that every $(j_n^{(1)})_z$ gives a $U(1)$ Chern-Simons gauge field, and combined with the gauge field dual to $(j_n^{(1)})_{\bar z}$, they form a $U(1)^\infty\times U(1)^\infty$ Chern-Simons gauge theory in the bulk.


\section{Approximately conserved higher spin currents}

The approximately conserved spin-1 current $(j_n^{(1)})_z$ generates a tower of approximately conserved higher spin currents, by the action of $W_N$ generators on $(j_n^{(1)})_z$. For example, $(j_1^{(1)})_z$ has a level-one $W$-descendent
\ie
(j_1^{(2)})_z&={1\over \sqrt{2(1-\lambda^2)}}\left(W^{(3)}_{-1}-{3\over 2}i\lambda L_{-1}\right)(j_1^{(1)})_z
\\
&=\sqrt{N\over 2\lambda^2(1-\lambda^2)}(W^{(3)}_{-2}-i\lambda\partial^2)\omega_1,
\fe
which is also a Virasoro primary\footnote{In appendix B, we fix the normalization of $(j^{(1)}_1)_z$ and check that it is a Virasoro primary.}. This is an approximately conserved stress tensor. The current non-conservation equation of $(j^{(1)}_1)_z$ then descends to the current non-conservation equation of $(j^{(2)}_1)_z$:
\ie\label{NCET}
\bar\partial (j_1^{(2)})_z&={1\over \sqrt{2(1-\lambda^2)}}\left(W^{(3)}_{-1}-{3\over 2}i\lambda L_{-1}\right)\bar\partial j_1^{(1)}
\\
&={i\lambda\over \sqrt{2N(1-\lambda^2)}}\left[(1-\lambda)\partial\phi_1\tilde\phi_1-(1+\lambda)\phi_1\partial\tilde\phi_1\right],
\fe
where we have used the null-state equations in appendix C. In general, the approximately conserved spin-1 current $(j_1^{(1)})_z$ has exactly one $W$-descendant Virasoro primary $(j^{(s)}_1)_z$ at each level $s$, which takes the form as
\ie\label{ACHSC}
(j^{(s)}_1)_z=\sqrt{N} (a_1W^{(s+1)}_{-s}+a_2\partial W^{(s)}_{-s+1}+\cdots+a_{s}\partial^{s})\omega_1,
\fe
where $a_i$ are some constants depending on $\lambda$, and can be fixed by requiring $(j^{(s)}_1)_z$ being a Virasoro primary. The $(j^{(s)}_1)_z$'s are approximately conserved higher spin currents. They satisfy the current non-conservation equations taking the form as
\ie\label{HSCNE}
\bar\partial (j^{(s)}_1)_z= {1\over \sqrt{N}}(b_1\partial^{s-1}\phi_1\tilde\phi_1+b_2\partial^{s-2}\phi_1\partial\tilde\phi_1+\cdots+b_s\phi_1\partial^{s-1}\tilde\phi_1),
\fe 
where $b_s$ are constants depending on $\lambda$, and can be fixed by requiring the left hand side of \eqref{HSCNE} being a Virasoro primary. By same argument, there are also antiholomorphic higher spin currents $(j^{(s)}_1)_{\bar z}$. We expect that there are also approximately conserved holomorphic and antiholomorphic higher spin currents $(j^{(s)}_2)_z$, $(j^{(s)}_3)_z$, and $(j^{(s)}_2)_{\bar z}$, $(j^{(s)}_3)_{\bar z}$ that take a the similar form as \eqref{ACHSC}.

\section{The single particle spectrum}

Now we state a conjecture on the complete spectrum of the single particle states in the bulk. Throughout this paper, by a single-trace operator we mean an operator that obeys the same large $N$ factorization property as single-trace operators in large $N$ gauge theories; such an operator is dual to the state of one elementary particle in the bulk. The products of single-trace operators are dual to multi-particle states. As we have seen in the previous section, the primary operators that involve up to one box in the Young tableaux of $\Lambda_+$ and $\Lambda_-$ are all single-trace operators: they are $\phi_1$, $\tilde\phi_1$, and $\omega_1$. The primaries that involve up to two boxes in the Young tableaux of $\Lambda_+$ and $\Lambda_-$ are some suitable linear combination of single-trace operators $\phi_2$, $\tilde\phi_2$, $\omega_2$, or products of two single-trace operators. We have also seen some evidences that the primaries with up to three boxes in their representations are  linear combinations of single-trace operators $\phi_3$, $\tilde\phi_3$, $\omega_3$, or products of single-trace operators. We conjecture that the primaries with up to $n$-box representations are linear combinations of single-trace operators $\phi_n$, $\tilde\phi_n$, $\omega_n$, or products of such single-trace operators $\phi_m, \tilde\phi_m, \omega_m$ for $m<n$. Here $\phi_n$ is a linear combination of primaries of the form $(\Lambda_+,\Lambda_-)$ that involve $(n,n-1)$ boxes, $\tilde\phi_n$ is a linear combination of primaries that involve $(n-1,n)$ boxes, and $\omega_n$ is a linear combination of light primaries of the form $(\Lambda,\Lambda)$ where $\Lambda$ involves $n$ boxes.

A part of this conjecture is easy to prove: the statement that there is only one light single-trace operator $\omega_n$ for each $n$ labeling the number of boxes in its corresponding $SU(N)$ representations follows easily from the fusion rule. First we note that generally, the light state of the form $(\Lambda,\Lambda)$ have dimension $B(\Lambda)\lambda^2/N+{\cal O}(N^{-2})$, where $B(\Lambda)$ is the number of boxes of the Young tableaux of the representation $\Lambda$, in the large $N$ limit and fixed finite $B(\Lambda)$. We may write a partition function of the light states
\ie
Z(x) = \sum_{(\Lambda,\Lambda)} x^{B(\Lambda)} = \prod_{n=1}^\infty {1\over 1-x^n}.
\fe
Each single-trace operator of dimension $n\lambda^2/N$ is a linear combination of $(\Lambda,\Lambda)$ with $B(\Lambda)=n$. The dimension of the product of single-trace operators is additive at order $1/N$. The products of a single-trace operator is counted by the partition function $1/(1-x^n)$. By comparing this with $Z(x)$, we see that there is precisely one single-trace operator $\omega_n$ for each $n$.

The $\phi_n$, $\tilde\phi_n$, $\omega_n$ are all the single-trace operators that are dual to scalar fields in the bulk. These are not all, however. There are other single-trace operators that are dual to spin-1, spin-2, and higher spin gauge fields. As explained in the previous section, while $\partial\omega_n$ is a level-one descendent of $\omega_n$, the norm of $\partial\omega_n$ goes to zero in the large $N$ limit. Consequently, the normalized operator $(j_n^{(1)})_z \sim \sqrt{N} \partial \omega_n$ behaves like a primary operator. Such operators will be referred to as {\it large $N$ primary operators}, and we include them in our list of single-trace operators because they should be dual to elementary fields in the bulk as well. We conjecture that $j_n^{(1)}$'s are single-trace operators dual to the spin-1 Chern-Simons gauge field in take bulk. This statement has passed some tests involving the three-point function of $j_n^{(1)}$ with two scalars. This is not the end of the story. As shown in the previous section, there are large $N$ primaries of higher spin $s$, denoted by $j^{(s)}_n$. These are single-trace operators dual to additional elementary higher spin gauge fields in the bulk. Unlike the original $W_N$ currents, however, the would-be higher spin symmetries generated by $j_n^{(s)}$ are broken by the boundary conditions on the charged scalars, leading to the current non-conservation relation. These hidden symmetries are recovered in the infinite $N$ limit.

Let us summarize our conjecture on the single-particle spectrum. There are two families of complex single-trace operators $\phi_n,\tilde\phi_n$, which are dual to massive complex scalar fields (of the same mass classically), one family of complex single-trace operators $\omega_n$, that are dual to massless scalars in the bulk, and a family of approximately conserved higher spin single-trace operators $j^{(s)}_n$ for each positive integer spin $s=1,2,3,\cdots$, that are dual to Chern-Simons spin-1 and higher spin gauge fields.


\section{Large $N$ partition functions}

In this section, we check our proposed single particle spectrum against the partition function of the $W_N$ minimal model and the $W_N$ characters of various primaries in the large $N$ limit.

Let $\cO$ be a single-trace operator with left and right dimensions $h_\cO$ and $\bar h_\cO$. From the bulk perspective, the (free) single particle contribution to the partition function is
\ie\label{CDOC}
& Z_{\cO}=\chi^\infty_\cO(h_\cO,q)\overline{\chi^\infty_\cO}(\bar h_\cO,\bar q),
\\
& \chi^\infty_\cO(h_\cO,q)={q^{h_\cO}\over 1-q} =q^{h_\cO}+q^{h_\cO+1}+q^{h_\cO+2}+\cdots.
\fe
The coefficients of the $q$-expansion are 1 in this case; they count the operators $\cO,\partial\cO,\partial^2\cO,\cdots$. There is an exception to the formula \eqref{CDOC}. If the single-trace operator $\cO$ has zero conformal weight, then the character is identity, i.e. $\chi^\infty_\cO(h_\cO=0)=1$. $\chi_{\cal O}^\infty$ is not the same as the $W_N$ character for the primary ${\cal O}$ in the large $N$ limit, because it misses the contribution from boundary excitations of higher spin fields. The contribution from  the boundary higher spin gauge fields to the partition function is
\ie
Z_{hs}=|\chi_{hs}^\infty|^2,~~~\chi_{hs}^\infty=\prod^\infty_{s=2}\prod^\infty_{n=s}{1\over (1-q^n)}.
\fe
According to our conjecture, the single particle partition functions is:
\ie\label{MSC}
&Z_{\phi_n}={q^{{1+\lambda\over 2}}\bar q^{{1+\lambda\over 2}}\over (1-q)(1-\bar q)},~~~Z_{\tilde\phi_n}={q^{{1-\lambda\over 2}}\bar q^{{1-\lambda\over 2}}\over (1-q)(1-\bar q)},
\fe
and
\ie
&Z_{\omega_n}=1,~~~Z_{j^{(s)}_n}=\chi^\infty_{j^{(s)}_n}={q^s\over 1-q}.
\fe
For simplicity, let us summarize all the character of higher spin gauge fields as a single character $\chi_{j_n}^\infty$:
\ie\label{S1CC}
\chi_{j_n}^\infty=\sum^\infty_{s=1}\chi^\infty_{j_n^{(s)}}=\sum_{s=1}^\infty{q^s\over 1-q}={q\over (1-q)^2}.
\fe

Next, let us consider the partition function for the $W_N$ minimal model in the large $N$ limit. Following from the diagonal modular invariance, the partition function in the large $N$ limit is given by the sum of the absolute value square of the characters:
\ie
Z_{W_N}=\sum_{(\Lambda_+,\Lambda_-)}|\chi_{(\Lambda_+,\Lambda_-)}|^2.
\fe
The characters $\chi_{(\Lambda_+,\Lambda_-)}$, for $\Lambda_\pm$ being representations with one to three boxes in the Young tableaux, in the large $N$ limit are computed in the appendix D up to cubic order. The following formulas in this section have all been checked up to this order. Let us start by looking at the contribution of the identity operator to the partition function, which in the large $N$ limit gives the partition function of the boundary higher spin gauge fields:
\ie
\lim_{N\to \infty}|\chi_{(0,0)}|^2=Z_{hs}.
\fe
The primary operators $(\yng(1),0)=\phi_1$ and $(0,\yng(1))=\tilde\phi_1$ are dual to massive scalars. Their contributions to the partition function indeed give the partition function of single massive scalar: 
\ie\label{DCPPH}
\lim_{N\to \infty}|\chi_{(\yng(1),0)}|^2= Z_{hs}Z_{\phi_1},
\\
\lim_{N\to \infty}|\chi_{(0,\yng(1))}|^2=Z_{hs}Z_{\tilde\phi_1}.
\fe
The primary operator $(\yng(1),\yng(1))=\omega_1$ is dual to a massless scalar, and its descendants $j^{(s)}_1$ are dual to spin-1, spin-2 and higher spin gauge fields. Also, by the current non-conservation equation \eqref{HSCNE}, some of the descendants of $(\yng(1),\yng(1))$ are dual to the two particle states. Therefore, contribution of $(\yng(1),\yng(1))$ to the partition function decomposes as
\ie
\lim_{N\to \infty}|\chi_{(\yng(1),\yng(1))}|^2=Z_{hs}(Z_{\omega_1}+\chi^\infty_{j_1}+\overline{\chi^\infty_{j_1}}+Z_{\phi_1}Z_{\tilde\phi_1}),
\fe
where the last term is the contribution of the two particle states of $\phi_1$ and $\tilde\phi_1$.

The identification of other primary operators are inevitable involving multi-particle states. By Bose statistics, we can write a multi-particle partition function in terms of the single-particle partition function \eqref{CDOC} as
\ie
Z^{multi}_\cO(t)&=\exp\left[\sum^\infty_{m=1}{Z_\cO(q^m,\bar q^m)\over m} t^m\right].
\fe
Suppose $\cO=\phi_n$, then the partition function $Z^{multi}_{\phi_n}(t)$ can be expanded as
\ie
Z^{multi}_{\phi_n}(t)=\sum_{\ell=0}^\infty t^\ell Z_{\phi_n^\ell},
\fe
where $Z_{\phi^m_n}$ is the $m$-particle partition function. For instance, $Z_{\phi_n^2}$ and $Z_{\phi_n^3}$ are given by
\ie
Z_{\phi_n^2}&
={q^{1+\lambda}\bar q^{1+\lambda}(1+q\bar q)\over (1- q)^2(1+ q)(1-\bar q)^2(1+\bar q)},
\\
Z_{\phi_n^3}&
={q^{{3\over 2}(1+\lambda)}\bar q^{{3\over 2}(1+\lambda)}(1+q\bar q+q^2\bar q+\bar q^2 q+q^2\bar q^2+q^3\bar q^3)\over (1- q)^3(1+ q)(1+ q+ q^2)(1-\bar q)^3(1+\bar q)(1+\bar q+\bar q^2)}.
\fe
For $\cO=\omega_n$, all the $m$-particle partition functions are identity:
\ie
Z_{\omega^m_n}
=1.
\fe
For $\cO=j^{(s)}_n$, the multi-particle partition function for $j^{(s)}_n$, $s=1,2,\cdots$, can be computed from
\ie
Z^{multi}_{j_n}(t)&=\prod^\infty_{s=1}Z^{multi}_{j^{(s)}_n}(t)=\exp\left[\sum^\infty_{m=1}\sum^\infty_{s=1}{\chi^\infty_{j^{(s)}_n}(q^m)\over m} t^m\right]=\exp\left[\sum^\infty_{m=1}{\chi^\infty_{j_n}(q^m)\over m} t^m\right].
\fe
Expanding $Z^{multi}_{j_n}(t)$ in powers of $t$, we can write the $Z^{multi}_{j_n}(t)$ as 
\ie
Z^{multi}_{j_n}(t)= 1+\chi^\infty_{j_n}t+\chi^\infty_{j_n^2}t^2+\chi^\infty_{j_n^3}t^3+\cdots,
\fe
where $\chi^\infty_{j_n^m}$ has the interpretation of the $m$-particle character (partition function). For instance,
\ie
\chi^\infty_{j_n^2}&={q^2(1+q^2)\over (1-q)^4(1+q)^2},
\\
\chi^\infty_{j_n^3}&={q^3(1+q^2+2q^2+q^4+q^6)\over (1-q)^6(1+q)^2(1+q+q^2)^2}.
\fe


Let us continue on the matching of boundary and bulk partition functions. Consider the primary operators $(\yng(1,1),0)$ and $(\yng(2),0)$. They are dual to two-particle states. 
Their contribution to the partition function matches with the two particle partition function:
\ie
\lim_{N\to \infty}\left(|\chi_{(\yng(2),0)}|^2+|\chi_{(\yng(1,1),0)}|^2\right)=Z_{hs}Z_{\phi_1^2}.
\fe
Now, consider the primary operators $(\yng(3),0)$, $(\yng(2,1),0)$, and $(\yng(1,1,1),0)$. They are dual to three-particle states. Their contribution to the partition function matches with the three-particle partition function:
\ie
\lim_{N\to \infty}\left(|\chi_{(\yng(3),0)}|^2+|\chi_{(\yng(2,1),0)}|^2+|\chi_{(\yng(1,1,1),0)}|^2\right)=Z_{hs}Z_{\phi_1^3}.
\fe
Next, consider the primary operators $(\yng(2),\yng(1))$ and $(\yng(1,1),\yng(1))$. Their contribution to the partition function also decomposes as the multi-particle partition functions:
\ie
\lim_{N\to \infty}\left(|\chi_{(\yng(2),\yng(1))}|^2+|\chi_{(\yng(2),\yng(1))}|^2\right)
=Z_{hs}\Big[Z_{\phi_1}\left(Z_{\omega_1}+\chi^\infty_{j_1}+\overline{\chi^\infty_{j_1}}\right)+Z_{\tilde\phi_1}Z_{\phi_1^2}+Z_{\phi_2}\Big].
\fe
For the primary operators $(\yng(2),\yng(2))$, $(\yng(1,1),\yng(1,1))$, $(\yng(2),\yng(1,1))$, and $(\yng(1,1),\yng(2))$, their contribution to the partition function decomposes as
\ie
&\lim_{N\to \infty}\left(|\chi_{(\yng(2),\yng(2))}|^2+|\chi_{(\yng(1,1),\yng(1,1))}|^2+|\chi_{(\yng(2),\yng(1,1))}|^2+|\chi_{(\yng(1,1),\yng(2))}|^2\right)
\\
&=Z_{hs}\Big[Z_{\omega_1^2}+Z_{\omega_1}(\chi^\infty_{j_1}+\overline{\chi^\infty_{j_1}})+(\chi^\infty_{j_1^2}+\overline{\chi^\infty_{j_1^2}})+|\chi^\infty_{j_1}|^2+Z_{\omega_1}Z_{\phi_1}Z_{\tilde\phi_1}
\\
&~~~+Z_{\phi_1}Z_{\tilde\phi_1}\left(\chi^\infty_{j_1}+\overline{\chi^\infty_{j_1}}\right)+Z_{\phi_1^2}Z_{\tilde\phi_1^2}+Z_{\omega_2}+(\chi^\infty_{j_2}+\overline{\chi^\infty_{j_2}})+Z_{\phi_2}Z_{\tilde\phi_1}+Z_{\phi_1}Z_{\tilde\phi_2}\Big].
\fe
Now, let us go on to the representations with three boxes in the Young tableaux. For the primary operators $(\yng(3),\yng(1))$, $(\yng(2,1),\yng(1))$, and $(\yng(1,1,1),\yng(1))$, their contribution to the partition function decomposes as 
\ie
&\lim_{N\to \infty}\Big(|\chi_{(\yng(3),\yng(1))}|^2+|\chi_{(\yng(2,1),\yng(1))}|^2+|\chi_{(\yng(1,1,1),\yng(1))}|^2\Big)
\\
&=Z_{hs}\Big[Z_{\phi_1}Z_{\phi_2}+\left(Z_{\omega_1}+\chi^\infty_{j_1}+\overline{\chi^\infty_{j_1}}\right)Z_{\phi_1^2}+Z_{\tilde\phi_1}Z_{\phi_1^3}\Big].
\fe
For the primary operators $(\yng(3),\yng(2))$, $(\yng(2,1),\yng(2))$, $(\yng(1,1,1),\yng(2))$, $(\yng(3),\yng(1,1))$, $(\yng(2,1),\yng(1,1))$, and $(\yng(1,1,1),\yng(1,1))$, their contribution to the partition function decomposes as 
\ie
&\lim_{N\to \infty}\Big(|\chi_{(\yng(3),\yng(2))}|^2+|\chi_{(\yng(2,1),\yng(2))}|^2+|\chi_{(\yng(1,1,1),\yng(2))}|^2+|\chi_{(\yng(3),\yng(1,1))}|^2+|\chi_{(\yng(2,1),\yng(1,1))}|^2+|\chi_{(\yng(1,1,1),\yng(1,1))}|^2\Big)
\\
&=Z_{hs}\Big[\left(Z_{\omega_2}+\chi^\infty_{j_2}+\overline{\chi^\infty_{j_2}}\right)Z_{\phi_1}+\left(Z_{\omega_1}+\chi^\infty_{j_1}+\overline{\chi^\infty_{j_1}}\right)Z_{\phi_2}+Z_{\phi_1^2}Z_{\tilde\phi_2}+Z_{\phi_1}Z_{\tilde\phi_1}Z_{\phi_2}
\\
&~~~~+\left(Z_{\omega_1^2}+Z_{\omega_1}\chi^\infty_{j_1}+Z_{\omega_1}\overline{\chi^\infty_{j_1}}+\chi^\infty_{j_1^2}+\overline{\chi^\infty_{j_1^2}}+\chi^\infty_{j_1}\overline{\chi^\infty_{j_1}}\right)Z_{\phi_1}
\\
&~~~~+\left(Z_{\omega_1}+\chi^\infty_{j_1}+\overline{\chi^\infty_{j_1}}\right)Z_{\phi_1^2}Z_{\tilde\phi_1}+Z_{\phi_1^3}Z_{\tilde\phi_1^2}+Z_{\phi_3}\Big].
\fe
The contribution from the primary operators $(\yng(3),\yng(3))$, $(\yng(2,1),\yng(3))$, $(\yng(1,1,1),\yng(3))$, $(\yng(3),\yng(2,1))$, $(\yng(2,1),\yng(2,1))$, $(\yng(1,1,1),\yng(2,1))$, $(\yng(3),\yng(1,1,1))$, $(\yng(2,1),\yng(1,1,1))$, and $(\yng(1,1,1),\yng(1,1,1))$,  to the partition function decomposes as 
\ie
&\lim_{N\to \infty}\Big(|\chi_{(\yng(3),\yng(3))}|^2+|\chi_{(\yng(2,1),\yng(3))}|^2+|\chi_{(\yng(1,1,1),\yng(3))}|^2+|\chi_{(\yng(3),\yng(2,1))}|^2
\\
&~~~+|\chi_{(\yng(2,1),\yng(2,1))}|^2+|\chi_{(\yng(1,1,1),\yng(2,1))}|^2+|\chi_{(\yng(3),\yng(1,1,1))}|^2+|\chi_{(\yng(2,1),\yng(1,1,1))}|^2+|\chi_{(\yng(1,1,1),\yng(1,1,1))}|^2\Big)
\\
&=Z_{hs}\Big[Z_{\omega^3_1}+Z_{\omega_1^2}\left(\chi^\infty_{j_1}+\overline{\chi^\infty_{j_1}}\right)+Z_{\omega_1}\left(\chi^\infty_{j_1^2}+\overline{\chi^\infty_{j_1^2}}+\chi^\infty_{j_1}\overline{\chi^\infty_{j_1}}\right)+\left(\chi^\infty_{j_1^3}+\overline{\chi^\infty_{j_1^3}}+\chi^\infty_{j_1^2}\overline{\chi^\infty_{j_1}}+\chi^\infty_{j_1}\overline{\chi^\infty_{j_1^2}}\right)
\\
&~~~~+\left(Z_{\omega^2_1}+Z_{\omega_1}\left(\chi^\infty_{j_1}+\overline{\chi^\infty_{j_1}}\right)+\left(\chi^\infty_{j_1^2}+\overline{\chi^\infty_{j_1^2}}+\chi^\infty_{j_1}\overline{\chi^\infty_{j_1}}\right)\right)Z_{\phi_1}Z_{\tilde\phi_1}+\left(Z_{\omega_1}+\left(\chi^\infty_{j_1}+\overline{\chi^\infty_{j_1}}\right)\right)Z_{\phi^2_1}Z_{\tilde\phi^2_1}+Z_{\phi^3_1}Z_{\tilde\phi^3_1}
\\
&~~~~+Z_{\omega_1}Z_{\omega_2}+Z_{\omega_1}\left(\chi^\infty_{j_2}+\overline{\chi^\infty_{j_2}}\right)+Z_{\omega_2}\left(\chi^\infty_{j_1}+\overline{\chi^\infty_{j_1}}\right)+\left(\chi^\infty_{j_1}+\overline{\chi^\infty_{j_1}}\right)\left(\chi^\infty_{j_2}+\overline{\chi^\infty_{j_2}}\right)
\\
&~~~~+\left(Z_{\omega_1}+\chi^\infty_{j_1}+\overline{\chi^\infty_{j_1}}\right)\left(Z_{\phi_1}Z_{\tilde\phi_2}+Z_{\phi_2}Z_{\tilde\phi_1}\right)+\left(Z_{\omega_2}+\chi^\infty_{j_2}+\overline{\chi^\infty_{j_2}}\right)Z_{\phi_1}Z_{\tilde\phi_1}+Z_{\phi^2_1}Z_{\tilde\phi_1}Z_{\tilde\phi_2}+Z_{\phi_1}Z_{\tilde\phi^2_1}Z_{\phi_2}
\\
&~~~~+Z_{\omega_1}+\chi^\infty_{j_1}+\overline{\chi^\infty_{j_1}}+Z_{\phi_1}Z_{\tilde\phi_3}+Z_{\phi_2}Z_{\tilde\phi_2}+Z_{\phi_3}Z_{\tilde\phi_1}\Big].
\fe

\section{Interactions and a semi-local bulk theory}

The three-point functions\footnote{Some three-point functions are computed, and a general form of such three-point functions are postulated in appendix E.} involving the hidden symmetry currents amount to the following assignment of gauge generators $T_n$ associated to the currents $j^{(s)}_n(z)$, which act on the states $|\phi_m\rangle$ and $|\tilde \phi_m\rangle$. We use the ket notation here, rather than the primary fields themselves, because while $\phi_m$ and $\tilde\phi_m$ have different scaling dimensions at infinite $N$, they are dual to scalar fields of the same mass that transform into one another under the hidden gauge symmetries.
\ie
&T_n|\phi_{m}\rangle = |\phi_{n+m}\rangle,~~~T_n|\bar\phi_m\rangle = -|\bar\phi_{m-n}\rangle~~(n<m)~~\text{or}~-|\tilde\phi_{n-m+1}\rangle~~(n\geq m),
\\
&T_n|\tilde\phi_{m}\rangle = - |\tilde\phi_{n+m}\rangle,~~~T_n|\bar{\tilde\phi}_m\rangle = |\bar{\tilde\phi}_{m-n}\rangle~~(n<m)~~\text{or}~|\phi_{n-m+1}\rangle~~(n\geq m).
\fe
Let us define the fields $\varphi_r$ and $\tilde\varphi_r$ for $r\in\bZ+{1\over 2}$ by
\ie
&\varphi_r=\phi_{r+{1\over 2}},~~~\varphi_{-r}=\bar{\tilde\phi}_{r+{1\over 2}},
\\
&\tilde\varphi_r=\tilde\phi_{r+{1\over 2}},~~~\tilde\varphi_{-r}=\bar{\phi}_{r+{1\over 2}}.
\fe
They are related by complex conjugation:
\ie
\bar\varphi_r=\tilde\varphi_{-r},~~~\bar{\tilde\varphi}_r=\varphi_{-r}.
\fe
In terms of $\varphi_r$ and $\tilde\varphi_r$, the gauge generators act as
\ie\label{crOPE}
&T_n|\varphi_r\rangle = |\varphi_{r+n}\rangle,~~~T_n|\tilde\varphi_r\rangle = -|\tilde\varphi_{r+n}\rangle.
\fe
We also have
\ie
&\overline T_n|\varphi_r\rangle = -|\varphi_{r-n}\rangle,~~~\overline T_n|\tilde\varphi_r\rangle=|\tilde\varphi_{r-n}\rangle.
\fe
which suggests the definition $T_{-n}=-\overline{T}_n$, or $j^{(s)}_{-n}=-\bar j^{(s)}_n$. Now \eqref{crOPE} is extended to all $n\in\bZ$. 
The action of $T_n$ can be diagonalized by the Fourier transform:
\ie
|\varphi(x)\rangle=\sum_{r\in\bZ+1/2} e^{irx}|\varphi_r\rangle,~~~|\tilde\varphi(x)\rangle=\sum_{r\in\bZ+1/2} e^{irx}|\tilde\varphi_r\rangle,~~~T(x)= \sum_{n\in\bZ}e^{inx}T_n,
\fe
where $x$ is an auxiliary generating parameter. Here we also included the generator $T_0$ which assigns charge $+1$ to $\varphi$ and charge $-1$ to $\bar\varphi$. With this definition, $|\bar\varphi(x)\rangle=|\tilde\varphi(x)\rangle,\overline T(x)=-T(x)$. We have
\ie
T(x)|\varphi(y)\rangle
=\delta(x-y)|\varphi(y)\rangle.
\fe
Here $x,y$ are understood to be periodically valued with periodicity $2\pi$.

What is the interpretation of this result? We see that there is a circle worth of gauge generators $T(x)$, each of which corresponds to a tower of gauge fields in $AdS_3$, of spin $s=1,2,3,\cdots,\infty$. Furthermore, these gauge generators commute, indicating Vasiliev theory with $U(1)^\infty$ ``Chan-Paton factor". At the level of bulk equation of motion, we expect the infinite family of Vasiliev theories to decouple. They only interact through the $AdS_3$ boundary conditions that mix the matter scalar fields. The boundary condition is such that the ``right moving" modes of $\varphi(x)$ on the circle, namely $\varphi_r$ with $r>0$ ($r={1\over 2}, {3\over 2},\cdots$) are dual to operators of dimension $\Delta_+=1+\lambda$, whereas $\varphi_r$ with $r<0$ are dual to operators of dimension $\Delta_- = 1-\lambda$. As a consequence of this boundary condition, the corresponding generating operator $\varphi(x;z,\bar z)$ in the CFT has two-point function
\ie
\vev{\varphi(x;z,\bar z)\bar\varphi(y;0)}&=\sum_{r,s\in\bZ+1/2}e^{irx+isy}\vev{\varphi_r(z)\tilde\varphi_s(0)}
=\left({1\over |z|^{2+2\lambda}}-{1\over |z|^{2-2\lambda}}\right) {i\over 2 \sin{x-y\over 2}}
\fe
in the large $N$ limit.

Note that the spin-1 gauge field is included here. It is also natural to include the massless scalar $\omega_n$, of spin $s=0$. $|\varphi(x)\rangle$ labels a complex massive scalar in $AdS_3$, for each $x$. This spectrum precisely fits into Vasiliev's system in three dimensions. In earlier works, we did not consider the spin-1 gauge field in Vasiliev theory, because it is governed by $U(1)\times U(1)$ Chern-Simons action and would decouple from the higher spin gravity if it weren't for the matter scalar field. It is possible to choose the boundary condition on the spin-1 Chern-Simons gauge field in $AdS_3$ so that there is no dual spin-1 current in the boundary CFT. This is presumably why the spin-1 current $j^{(1)}_0(z)$ is missing from the spectrum of $W_N$ minimal model. But the spin-1 currents $j^{(1)}_n(z)$ do exist in the infinite $N$ limit. Usually, in three-dimensional Vasiliev theory, there is no propagating massless scalar field either. There is however, an auxiliary scalar field $C_{aux}$ \cite{Chang:2011mz}, whose equation of motion at the linearized level takes the form $\nabla_\mu C_{aux}=0$. Classically, we could trade this equation with the massless Klein-Gordon equation $\Box C_{aux}=0$, together with the $\Delta=0$ boundary condition which eliminates normalizable finite energy states of this field in $AdS_3$. If this scalar field acquires a small mass, of order $1/N$ due to quantum corrections, then the boundary condition would allow for a normalizable state in $AdS_3$ of very small energy/conformal weight. We believe that this is the origin of the elementary light scalars $\omega_n$ themselves, in the infinite family of Vasiliev systems parameterized by the circle.

The identification of the single-trace operators, dual to elementary particles in the bulk, makes sense a priori only in the infinite $N$ limit. Non-perturbatively, or at finite $N,k$, the infinite family $\phi_n, \tilde\phi_n, \omega_n, j_n^{(s)}$ should be cut off to a finite family. Due to the restrictions on the unitary representations of $SU(N)$ current algebra at level $k$ or $k+1$, we expect the subscript $n$ which counts the number of boxes in the Young tableau in the construction of the single-trace primaries to be cut off at $n\sim k$. This means that the circle that parameterize a continuous family of Vasiliev theories in $AdS_3$ should be rendered discrete, with spacing $\sim 2\pi/k$.


\section{Discussion}

We have proposed that the holographic dual of $W_N$ minimal model in the 't Hooft limit, $k,N\to \infty$, $0<\lambda<1$, is a circle worth of Vasiliev theories in $AdS_3$ that couple with one another only through the boundary conditions on the matter scalars, which break all but one single tower of higher spin symmetries. It would seem to be a natural question to ask what is the CFT dual to the bulk theory with symmetry-preserving boundary conditions, that assign say the same scaling dimension $\Delta_+$ to all matter scalars. If we are to flip the boundary condition on $\tilde\phi_n$, on the CFT side this corresponds to turning on the double trace deformation by $\tilde\phi_n \bar{\tilde\phi}_n$ and flow to the critical point (IR in this case). This deformation decreases the central charge $c\approx N(1-\lambda^2)$ by an order $N^0$ amount. It is unclear what is the fixed point one ends up with by turning on double trace deformations $\tilde\phi_n \bar{\tilde\phi}_n$ for all $n$ (which should be cut off at $\sim k$), if there is such a nontrivial critical point at all.

There has been an alternative proposal on the holographic dual of $W_N$ minimal model \cite{Castro:2011iw,Perlmutter:2012ds,Gaberdiel:2012ku}, as Vasiliev theory based on $hs[N]\simeq sl(N)$ higher spin algebra, with families of conical deficit solutions included to account for the primaries missing from the perturbative spectrum of Vasiliev theory. On the face of it, this proposal involves an entirely different limit, where $N$ is held fixed, and an analytic continuation is performed in $k$ so that the central charge $c$ is large. The resulting CFT is not unitary. Furthermore, it is unclear to us that the analog of large $N$ (or rather, large $c$) factorization holds in this limit, which would be necessary for the holographic dual to be weakly coupled.

There is also an intriguing parallel between the 't Hooft limit of $W_N$ minimal model in two dimensions and Chern-Simons vector model in three dimensions. While the gauge invariant local operators and their correlation functions on $\mathbb{R}^3$ or $S^3$ in the three dimensional Chern-Simons vector model are expected to be computed by the parity violating Vasiliev theory in $AdS_4$ to all order in $1/N$, the duality in its naive form is not expected to hold for the CFT on three-manifolds of nontrivial topology (e.g. when the spatial manifold is a torus or a higher genus surface). This is because the topological degrees of freedom of the Chern-Simons gauge fields cannot be captured by a semi-classical theory in the bulk with Newton's constant that scales like $1/N$ rather than $1/N^2$. In a similar manner, the $W_N$ minimal model CFT on $\mathbb{R}^2$ or $S^2$ in the large $N$ admits a closed subsector, generated by the OPEs of the primary $\phi_1$ along with higher spin currents, that is conjectured to be perturbatively dual to Vasiliev theory in $AdS_3$. This duality makes sense only perturbatively in $1/N$. The light primaries which in a sense arise from twistor sectors must be included to ensure that the CFT is modular invariant. Here we see that the bulk theory should be extended as well, to an infinite family of Vasiliev theories. It would be interesting to understand the analogous statement in the $AdS_4/CFT_3$ example, where the connection to ordinary string theory is better understood \cite{Chang:2012kt} .

\bigskip

\section*{Acknowledgments}

We would like to thank Matthias Gaberdiel, Rajesh Gopakumar, Tom Hartman, Shiraz Minwalla, Soo-Jong Rey and Steve Shenker for useful discussions. We would like to especially thank the hospitality of the 6th Asian Winter School on Strings, Particles, and Cosmology in Kusatsu, Japan, 2012 Indian Strings Meeting, Puri, India, and Tata Institute for Fundamental Research, Mumbai, India, during the course of this work. XY would like to thank the organizers of CQUeST Spring Workshop on Higher Spins and String Geometry, Seoul, Ginzburg conference at Lebedev Institute, Moscow,  Strings 2012, Munich, and Komaba 2013, Tokyo, where partial results of this work are presented. 
This work is supported in part by the Fundamental Laws Initiative 
Fund at Harvard University, and by NSF Award PHY-0847457.

\appendix
\section{Higher spin charges}
The higher spin charges of primary operators can be computed using the Coulomb gas formalism reviewed in \cite{Bouwknegt:1992wg,Bilal:1991eu,Chang:2011vka}. The higher spin currents $W^{(s)}$ can be constructed by first considering the differential operator of order $N$:
\ie\label{UUUUUUUU}
(2iv_0)^N\cD_N&=:\prod^N_{i=1}(2iv_0\partial+h_i\cdot\partial X):,
\fe
then expanding this differential operator as
\ie
\cD_N=\partial^N+\sum^N_{s=1}(2iv_0)^{-k}U^{(s)}\partial^{N-s}.
\fe
Expanding the right hand side of \eqref{UUUUUUUU}, we obtain $U^{(1)}=0$ and $U^{(2)}=-{1\over 2}:\partial X\cdot \partial X:+2v_0\rho\cdot\partial^2 X$, which is the stress tensor. For $s>2$, $U^{(s)}$ behaves like a spin-$s$ higher spin current, but it is not primary. The dimension-$s$ primary operator $W^{(s)}$ can be constructed from $U^{(s)}$, for example \cite{DiFrancesco:1990qr}:
\ie\label{UtoW}
&W^{(2)}=U^{(2)},
\\
&W^{(3)}=U^{(3)}-{N-2\over 2}(2iv_0)\partial U^{(2)},
\\
&W^{(4)}=U^{(4)}-{N-3\over 2}(2iv_0)\partial U^{(3)}+{(N-2)(N-3)\over 10}(2iv_0)^2\partial^2 U^{(2)}
\\
&~~~~~~~~~-{(N-2)(N-3)(5N+7)\over 10N^2(N^2-1)}:(U^{(2)})^2:,
\\
&W^{(5)}=U^{(5)}-{N-4\over 2}(2iv_0)\partial U^{(4)}+{3(N-3)(N-4)\over 28}(2iv_0)^2\partial^2U^{(3)}
\\
&~~~~~~~~~-{(N-2)(N-3)(N-4)\over 84}(2iv_0)^3\partial^3U^{(2)}
\\
&~~~~~~~~+{(N-3)(N-4)(7N+13)\over 14N(N^2-1)}\Big((N-2)(2iv_0):U^{(2)}\partial U^{(2)}:-2:U^{(3)}U^{(2)}:\Big).
\fe
The the zero mode of the quasi-primaries $U^{(s)}$ acting on a primary operator $\exp(iv\cdot X)$ gives
\ie
u_k(v)=(-i)^{k-1}\sum_{i_1<\cdots<i_k}\prod^k_{j=1} \left(v\cdot h_{i_j}+(k-j)v_0\right).
\fe
where
\ie
v=\sqrt{p'\over p}\Lambda_+ -\sqrt{p\over p'}\Lambda_-,~~~~h_k=\omega_1-\sum^{k-1}_{i=1}\A_{i}.
\fe
The higher spin charges $w_k$ are by $u_k$ from \eqref{UtoW}. We conjecture that the higher spin charges of $\phi_n$, in the large $N$ limit, are
\ie
w_s(\phi_n)={i^{s-2}\Gamma(s)^2\Gamma(\lambda+s)\over \Gamma(2s-1)\Gamma(1+\lambda)},
\fe
and the higher spin charges of $\tilde\phi_n$ are
\ie
w_s(\tilde\phi_n)={(-i)^{s-2}\Gamma(s)^2\Gamma(\lambda+s)\over \Gamma(2s-1)\Gamma(1+\lambda)}.
\fe
These two formulas are checked up to $n=2,s=5$. We also conjecture that in the large $N$ limit the higher spin charges of $\omega_n$ are $n$ times the higher spin charges of $\omega_1$, and the higher spin charges of $\omega_n$ up to spin-5 are
\ie
&w_2(\omega_n)={n\lambda^2\over 2N},
\\
&w_3(\omega_n)=i{n\lambda^3\over 6N},
\\
&w_4(\omega_n)=-{n\lambda^2(1+\lambda^2)\over 20N},
\\
&w_5(\omega_n)=-i{n\lambda^3(5+\lambda^2)\over 70N}.
\fe
The above formulas are checked up to $n=3$.

\section{An approximately conserved spin-2 current}
The approximately conserved spin-2 field takes the form as
\ie
(j^{(2)}_1)_z=\A\left(W^{(3)}_{-1}-{3\over 2}i\lambda L_{-1}\right)L_{-1}\omega_1=\A (W^{(3)}_{-2}-i\lambda\partial^2)\omega_1,
\fe
where $\A$ is a normalization constant. We check that this is a Virasoro primary operator:
\ie 
&L_1(W^{(3)}_{-2}-i\lambda\partial^2)\omega_1=\left[4W^{(3)}_{-1}-2i\lambda L_{-1}\right]\omega_1=0,
\\
&L_{2}(W^{(3)}_{-2}-i\lambda\partial^2)\omega_1=(6w_3-6ih\lambda)\omega_1=0,
\fe
where we have used the null-state equation for $\omega_1$:
\ie
W^{(3)}_{-1}\omega_1={i\lambda\over 2}\partial\omega_1.
\fe
Let us compute the normalization constant $\A$. Considering the three-point function $\vev{W(z)\bar\omega_1(z_1)(W^{(3)}_{-2}-i\lambda\partial^2)\omega_1(z_2)}$, since it is a three-point function of three conformal primaries, it takes the form as
\ie\label{TPF}
\vev{W(z)\bar\omega_1(z_1)(W^{(3)}_{-2}-i\lambda\partial^2)\omega_1(z_2)}={a_1\over (z-z_1)(z-z_2)^5(z_1-z_2)^{-1}}.
\fe
The structure constant $a_1$ can be determined by performing contour integral $\oint_{z_2} dz(z-z_2)^4$ on the both hand side. On RHS, we obtain
\ie
\oint_{z_2} dz{a_1\over (z-z_1)(z-z_2)(z_1-z_2)^{-1}}=-a_1
\fe
On LHS, we have
\ie
\vev{\bar\omega_1(z_1)W^{(3)}_{2}(W^{(3)}_{-2}-i\lambda\partial^2)\omega_1(z_2)}&=\vev{\bar\omega_1(z_1)\left(8W^{(4)}_0+{4\over 5}(\lambda^2-4)L_0-12i\lambda W^{(3)}_0\right)\omega_1(z_2)}
\\
&=-{2\lambda^2(1-\lambda^2)\over N}.
\fe
Now, we perform a contour integral $\int_{z_1}dz$ on \eqref{TPF}, we obtain
\ie
\vev{W^{(3)}_{-2}\bar\omega_1(z_1)(W^{(3)}_{-2}-i\lambda\partial^2)\omega_1(z_2)}&=\oint_{z_1}dz{a_1\over (z-z_1)(z-z_2)^5(z_1-z_2)^{-1}}
\\
&={a_1\over (z_1-z_2)^4}={2\lambda^2(1-\lambda^2)\over N}{1\over (z_1-z_2)^4}.
\fe
Using similar method, we obtain
\ie
\vev{\bar\omega_1(z_1)W^{(3)}_{-2}\omega_1(z_2)}&=\oint_{z_2}dz{b\over (z-z_1)^3(z-z_2)^3(z_1-z_2)^{-3}}
\\
&={6b\over (z_2-z_1)^5(z_1-z_2)^{-3}}
={i\lambda^3\over N}{1\over (z_1-z_2)^2}.
\fe
We have
\ie
\vev{(W^{(3)}_{-2}+i\lambda\partial^2)\bar\omega_1(z_1)(W^{(3)}_{-2}-i\lambda\partial^2)\omega_1(z_2)}={2\lambda^2(1-\lambda^2)\over N}{1\over (z_1-z_2)^4}.
\fe
The normalization constant $\A$ is
\ie
\A=\sqrt{N\over 2\lambda^2(1-\lambda^2)}.
\fe

\section{Null-state equations}

The $W_{\infty}[\lambda]$, in the $c\to \infty$ limit, and truncating to the generators $W^{(s)}_{n}$, $|n|<s$, reduces to the wedge algebra $hs(\lambda)$, which is given by
\ie\label{WAC}
&[W^{(s)}_{m},W^{(t)}_n]=\sum^{s+t-|s-t|-1}_{u=2,4,6,\cdots}g^{st}_u(m,n;\lambda)W^{(s+t-u)}_{m+n},
\fe
where the structure constant $g^{st}_u(m,n;\lambda)$ is
\ie
g^{st}_u(m,n;\lambda)={q^{u-2}\over 2(u-1)!}\phi^{st}_u(\lambda) N^{st}(m,n),
\fe
and
\ie
N^{st}_u(\lambda)=\sum_{k=0}^{u-1} {(-1)^k\Gamma(u)\Gamma(s-m)\Gamma(s+m)\Gamma(t-n)\Gamma(t+n)\over \Gamma(1+k)\Gamma(u-k)\Gamma(s-m-k)\Gamma(s+m+k-u+1)\Gamma(t+n-k)\Gamma(t-n+k-u+1)},
\fe
and
\ie
\phi^{st}_u(\lambda)={}_4F_3\left[\begin{matrix}
 {1\over 2}  + \lambda ~,~    {1\over 2}  - \lambda  ~,~ {2-u\over 2} ~ ,~ {1-u\over 2}
 \\
 {3\over 2}-s ~ , ~~ {3\over 2} -t~ ,~~ {1\over 2} + s+t-u
\end{matrix}\Bigg|1\right].
\fe
$q$ is an arbitrary constant controls the normalization of the higher spin generators. In our convention, $q=i/4$. Using this commutator \eqref{WAC}, we can derived a set of null-state equations for $\phi_n$, $\tilde\phi_n$ and $\omega_n$. 

Consider a primary operator $\cO$. The descendants of $\cO$ can be separated into two classes. The descendants in the first class are the operators take the form as a combination of $W^{(s)}_{-n}$, $0<n<s$, acting on $\cO$. The rest of the descendants are in the second class. The descendants in the first class have the norm of order one, and the descendants in the second class have the norm of order $N$. The bulk dual of the descendants in the first class are single- or multi-particle states without boundary higher spin gauge field excitation, and the bulk dual descendants in the second class are the states with boundary higher spin gauge field excitations. Now, let us focus on the primary $\phi_1$. The partition of $\phi_1$ takes the form as \eqref{DCPPH} after modding out the boundary higher spin gauge field character $\chi_{hs}^\infty$, which means that at each level, there is only one independent descendent in the first class, which are $\partial^m\phi_1$'s. Therefore, the Kac matrices
\ie
\begin{pmatrix}
[L_{1}^n,L_{-1}^n],[W^{(s)}_n,L_{-1}^n]
\\
[L_1^n,W^{(s)}_{-n}],[W^{(s)}_n,W^{(s)}_{-n}]
\end{pmatrix},
\fe
for $0<n<s$, are rank 1, and have a singular vector, which gives the null-state equation:
\ie
&W^{(s)}_{-n}\phi_1
={i^{s-2}\Gamma(s)\Gamma(n+s)\Gamma(s+\lambda)\over\Gamma(n+1)\Gamma(2s-1)\Gamma(n+\lambda+1)}\partial^n\phi_1.
\fe
Similarly, we have the null-state equation for $\tilde\phi_1$:
\ie
&W^{(s)}_{-n}\tilde\phi_1={(-i)^{s-2}\Gamma(s)\Gamma(n+s)\Gamma(s-\lambda)\over\Gamma(n+1)\Gamma(2s-1)\Gamma(n-\lambda+1)}\partial^n\tilde\phi_1.
\fe

Next, let us consider the operator $\omega_1$. After moving out the character of boundary higher spin gauge fields, the character of $\omega_1$ takes the form as
\ie
\chi^\infty_{\omega_1}+\chi^\infty_{j_1}=1+q+2q^2+3q^3+\cdots.
\fe
At level one, there is one descendent in the first class, which is $(j^{(1)}_1)_z$ or $\partial\omega_1$, and the Kac matrix
\ie
\begin{pmatrix}
[L_{1},L_{-1}],[W^{(s)}_{1},L_{-1}]
\\
[L_1,W^{(s)}_{-1}],[W^{(s)}_1,W^{(s)}_{-1}]
\end{pmatrix},
\fe
is rank one, and gives the null-state equations:
\ie
W_{-1}^{(s)}\omega_1={sw_s\over 2h}\partial\omega_1.
\fe
Plugging in the value of the higher spin charges, we obtain
\ie
&W_{-1}^{(3)}\omega_1=i{\lambda\over 2}\partial\omega_1,
\\
&W_{-1}^{(4)}\omega_1=-{1+\lambda^2\over 5}\partial\omega_1,
\\
&W_{-1}^{(5)}\omega_1=-i{\lambda(5+\lambda^2)\over 14}\partial\omega_1.
\fe
At level two, there are two descendants in the first class. They are $\partial j^{(1)}_1$ and $j^{(2)}_1$ or $\partial^2\omega_1$ and $W^{(3)}_{-2}\omega_1$. The Kac matrix
\ie
\begin{pmatrix}
[L^2_{1},L^2_{-1}],[W^{(3)}_{2},L^2_{-1}],[W^{(4)}_{2},L^2_{-1}]
\\
[L^2_1,W^{(3)}_{-2}],[W^{(3)}_2,W^{(3)}_{-2}],[W^{(s)}_2,W^{(3)}_{-2}]
\\
[L^2_1,W^{(s)}_{-2}],[W^{(3)}_2,W^{(s)}_{-2}],[W^{(s)}_2,W^{(s)}_{-2}]
\end{pmatrix},
\fe
has one singular vector. For $s=4$, this gives the null state equation
\ie
W^{(4)}_{-2}\omega_1&=-{1\over 2}\partial^2\omega_1+i{\lambda\over 2}W^{(3)}_{-2}\omega_1.
\fe
In general, at level $n$, there are $n$ independent descendants in the first class, and they are  $\partial^{n-1} j^{(1)}_1$, $\partial^{n-2}j^{(2)}_1$, $\cdots$, and $j^{(n)}_1$, or $\partial^n\omega_1$, $\partial^{n-1}W^{(3)}_{-2}\omega_1$, $\cdots$, and $W^{(n+1)}_{-n}\omega_1$, where we have suppressed the $z$ index.

\section{$W_N$ characters}

As reviewed in \cite{Bouwknegt:1992wg,Chang:2011vka}, the characters of the primary operators in the $W_N$ minimal model are given by the formula
\ie\label{CHWN}
\chi_{(\Lambda_+,\Lambda_-)}={1\over \eta(\tau)^{N-1}}\sum_{w\in W,n\in \Gamma_{pp'}}\epsilon(w)q^{{1\over 2}|w(\lambda)+\lambda'+n|^2+{c\over 24}}
\fe
where $p=k+N$, $p'=k+N+1$, $W$ is the Weyl group, $\Gamma_{pp'}$ is $\sqrt{pp'}$ times the root lattice $\Lambda_{root}$, and $\lambda$, $\lambda'$ are
\ie
\lambda=\sqrt{p'\over p}(\Lambda_++\rho),~~~\lambda'=-\sqrt{p\over p'}(\Lambda_-+\rho).
\fe
In the large $N$ limit, the terms with nonzero $n$ in the summation over the lattice $\Gamma_{pp'}$ are of order $\cO(q^{N})$, and can be ignored. By evaluating the formula \eqref{CHWN}, we obtain the following characters:
\ie
&\chi_{(0,0)}=1+q^2+2q^3+\cdots
\\
&\chi_{(\yng(1),0)}=q^{h_{\Yboxdim{3pt}(\yng(1),0)}}\left(1+q+2q^2+4q^3+\cdots\right)
\\
&\chi_{(\yng(1,1),0)}=q^{h_{\Yboxdim{3pt}(\yng(1,1),0)}}\left(1+q+3q^2+5q^3+\cdots\right)
\\
&\chi_{(\yng(2),0)}=q^{h_{\Yboxdim{3pt}(\yng(2),0)}}\left(1+q+3q^2+5q^3+\cdots\right)
\\
&\chi_{(\yng(1,1,1),0)}=q^{h_{\Yboxdim{3pt}(\yng(1,1,1),0)}}\left(1+q+3q^2+6q^3+\cdots\right)
\\
&\chi_{(\yng(2,1),0)}=q^{h_{\Yboxdim{3pt}(\yng(2,1),0)}}\left(1+2q+4q^2+9q^3+\cdots\right)
\\
&\chi_{(\yng(3),0)}=q^{h_{\Yboxdim{3pt}(\yng(3),0)}}\left(1+q+3q^2+6q^3+\cdots\right)
\\
&\chi_{(\yng(1),\yng(1))}=q^{h_{\Yboxdim{3pt}(\yng(1),\yng(1))}}\left(1+q+3q^2+6q^3+\cdots\right)
\\
&\chi_{(\yng(2),\yng(2))}=q^{h_{\Yboxdim{3pt}(\yng(2),\yng(2))}}\left(1+q+3q^2+6q^3+\cdots\right)
\\
&\chi_{(\yng(1,1),\yng(1,1))}=q^{h_{\Yboxdim{3pt}(\yng(1,1),\yng(1,1))}}\left(1+q+4q^2+8q^3+\cdots\right)
\\
&\chi_{(\yng(3),\yng(3))}=q^{h_{\Yboxdim{3pt}(\yng(3),\yng(3))}}\left(1+q+3q^2+6q^3+\cdots\right)
\\
&\chi_{(\yng(2,1),\yng(2,1))}=q^{h_{\Yboxdim{3pt}(\yng(2,1),\yng(2,1))}}\left(1+2q+6q^2+14q^3+\cdots\right)
\\
&\chi_{(\yng(1,1,1),\yng(1,1,1))}=q^{h_{\Yboxdim{3pt}(\yng(1,1,1),\yng(1,1,1))}}\left(1+q+4q^2+9q^3+\cdots\right)
\\
&\chi_{(\yng(2),\yng(1))}=q^{h_{\Yboxdim{3pt}(\yng(2),\yng(1))}}\left(1+q+3q^2+6q^3+\cdots\right)
\\
&\chi_{(\yng(1,1),\yng(1))}=q^{h_{\Yboxdim{3pt}\Yvcentermath1(\yng(1,1),\yng(1))}}\left(1+2q+4q^2+9q^3+\cdots\right)
\\
&\chi_{(\yng(3),\yng(2))}=q^{h_{\Yboxdim{3pt}(\yng(3),\yng(2))}}\left(1+q+3q^2+6q^3+\cdots\right)
\\
&\chi_{(\yng(2,1),\yng(2))}=q^{h_{\Yboxdim{3pt}(\yng(2,1),\yng(2))}}\left(1+2q+5q^2+11q^3+\cdots\right)
\\
&\chi_{(\yng(2,1),\yng(1,1))}=q^{h_{\Yboxdim{3pt}(\yng(2,1),\yng(1,1))}}\left(1+2q+5q^2+12q^3+\cdots\right)
\\
&\chi_{(\yng(1,1,1),\yng(1,1))}=q^{h_{\Yboxdim{3pt}(\yng(1,1,1),\yng(1,1))}}\left(1+2q+5q^2+11q^3+\cdots\right)
\\
&\chi_{(\yng(2,1),\yng(1))}=q^{h_{\Yboxdim{3pt}(\yng(2,1),\yng(1))}}\left(1+2q+5q^2+11q^3+\cdots\right)
\\
&\chi_{(\yng(1,1,1),\yng(1))}=q^{h_{\Yboxdim{3pt}(\yng(1,1,1),\yng(1))}}\left(1+2q+5q^2+10q^3+\cdots\right)
\\
&\chi_{(\yng(3),\yng(1))}=q^{h_{\Yboxdim{3pt}(\yng(3),\yng(1))}}\left(1+q+3q^2+6q^3+\cdots\right)
\fe
\ie
&\chi_{(\yng(2),\yng(1,1))}=q^{h_{\Yboxdim{3pt}(\yng(2),\yng(1,1))}}\left(1+2q+5q^2+10q^3+\cdots\right)
\\
&\chi_{(\yng(1,1),\yng(2))}=q^{h_{\Yboxdim{3pt}(\yng(1,1),\yng(2))}}\left(1+2q+5q^2+10q^3+\cdots\right)
\\
&\chi_{(\yng(1,1,1),\yng(2))}=q^{h_{\Yboxdim{3pt}(\yng(1,1,1),\yng(2))}}\left(1+2q+6q^2+12q^3+\cdots\right)
\\
&\chi_{(\yng(3),\yng(1,1))}=q^{h_{\Yboxdim{3pt}(\yng(3),\yng(1,1))}}\left(1+2q+5q^2+11q^3+\cdots\right)
\\
&\chi_{(\yng(3),\yng(2,1))}=q^{h_{\Yboxdim{3pt}(\yng(3),\yng(2,1))}}\left(1+2q+5q^2+11q^3+\cdots\right)
\\
&\chi_{(\yng(2,1),\yng(1,1,1))}=q^{h_{\Yboxdim{3pt}(\yng(2,1),\yng(1,1,1))}}\left(1+3q+7q^2+17q^3+\cdots\right)
\\
&\chi_{(\yng(3),\yng(1,1,1))}=q^{h_{\Yboxdim{3pt}(\yng(3),\yng(1,1,1))}}\left(1+2q+6q^2+13q^3+\cdots\right)
\\
&\chi_{(\yng(2,1),\yng(3))}=q^{h_{\Yboxdim{3pt}(\yng(2,1),\yng(3))}}\left(1+2q+5q^2+11q^3+\cdots\right)
\\
&\chi_{(\yng(1,1,1),\yng(2,1))}=q^{h_{\Yboxdim{3pt}(\yng(1,1,1),\yng(2,1))}}\left(1+3q+7q^2+17q^3+\cdots\right)
\\
&\chi_{(\yng(1,1,1),\yng(3))}=q^{h_{\Yboxdim{3pt}(\yng(1,1,1),\yng(3))}}\left(1+2q+6q^2+13q^3+\cdots\right)
\fe

\section{Some three-point functions}

In this section, we will compute several three-point functions involving the approximately conserved spin-1 current $(j_n^{(1)})_z$ in the large $N$ limit. For simplicity, we will suppress the index $z$ in the following discussion. Let us first consider the three-point functions of the form $\vev{j_n^{(1)}\bar\phi_m\bar{\tilde\phi}_{n-m+1}}$. They are given by taking a derivative on the three-point function $\vev{\omega_n\bar\phi_m\bar{\tilde\phi}_{n-m+1}}$. For example, by taking one derivative on
\ie
&\vev{\omega_1(z_1)\bar\phi_1(z_2)\bar{\tilde\phi}_1(z_3)}={1\over N}{1\over |z_{12}|^{2\lambda}|z_{23}|^2|z_{13}|^{-2\lambda}},
\fe
we obtian
\ie
&\vev{j_1^{(1)}(z_1)\bar\phi_1(z_2)\bar{\tilde\phi}_1(z_3)}={1\over \sqrt{N}}{1\over |z_{12}|^{2\lambda}|z_{23}|^2|z_{13}|^{-2\lambda}}\left({1\over z_{13}}-{1\over z_{12}}\right).
\fe
Similarly by taking a derivative on \eqref{OPhPh2} and \eqref{OPhPh2}, we obtain
\ie
&\vev{j^{(1)}_2(z_1)\bar\phi_1(z_2)\bar{\tilde\phi}_2(z_3)}=\vev{j^{(1)}_2(z_1)\bar\phi_2(z_2)\bar{\tilde\phi}_1(z_3)}=\vev{j^{(1)}_3(z_1)\bar\phi_1(z_2)\bar{\tilde\phi}_3(z_3)}=\vev{j^{(1)}_3(z_1)\bar\phi_3(z_2)\bar{\tilde\phi}_1(z_3)}
\\
&={1\over \sqrt{N}}{1\over |z_{12}|^{2\lambda}|z_{23}|^2|z_{13}|^{-2\lambda}}\left({1\over z_{13}}-{1\over z_{12}}\right).
\fe
We postulate the general form of the three-point function to be
\ie
\vev{j^{(1)}_n(z_1)\bar\phi_m(z_2)\bar{\tilde\phi}_{n-m+1}(z_3)}={1\over \sqrt{N}}{1\over |z_{12}|^{2\lambda}|z_{23}|^2|z_{13}|^{-2\lambda}}\left({1\over z_{13}}-{1\over z_{12}}\right).
\fe
Next, let us consider the three-point function of the form $\vev{j_n^{(1)}\phi_m\bar\phi_{n+m}}$ and $\vev{j_n^{(1)}\tilde\phi_m\bar{\tilde\phi}_{n+m}}$. The computation of this three-point function is a bit subtle. Let us first show an example $\vev{  j^{(1)}_1(z_1)\phi_1(z_2)\bar\phi_2(z_3)}$. To compute this three-point function, we consider the three-point functions:
\ie
&\vev{\omega_1(z_1)\phi_1(z_2)(\overline{\yng(2)},\overline{\yng(1)})(z_3)}={1\over \sqrt{2}}{1\over |z_{23}|^{2h_{\Yboxdim{3pt}(\yng(1),0)}+2h_{\Yboxdim{3pt}(\overline{\yng(2)},\overline{\yng(1)})}-2h_{\Yboxdim{3pt}(\yng(1),\yng(1))}}|z_{12}|^{2h_{\Yboxdim{3pt}(\yng(1),0)}+2h_{\Yboxdim{3pt}(\yng(1),\yng(1))}-2h_{\Yboxdim{3pt}(\overline{\yng(2)},\overline{\yng(1)})}}|z_{13}|^{2h_{\Yboxdim{3pt}(\overline{\yng(2)},\overline{\yng(1)})}+2h_{\Yboxdim{3pt}(\yng(1),\yng(1))}-2h_{\Yboxdim{3pt}(\yng(1),0)}}},
\\
&\vev{\omega_1(z_1)\phi_1(z_2)(\overline{\yng(1,1)},\overline{\yng(1)})(z_3)}={1\over \sqrt{2}}{1\over |z_{23}|^{2h_{\Yboxdim{3pt}(\yng(1),0)}+2h_{\Yboxdim{3pt}(\overline{\yng(1,1)},\overline{\yng(1)})}-2h_{\Yboxdim{3pt}(\yng(1),\yng(1))}}|z_{12}|^{2h_{\Yboxdim{3pt}(\yng(1),0)}+2h_{\Yboxdim{3pt}(\yng(1),\yng(1))}-2h_{\Yboxdim{3pt}(\overline{\yng(1,1)},\overline{\yng(1)})}}|z_{13}|^{2h_{\Yboxdim{3pt}(\overline{\yng(1,1)},\overline{\yng(1)})}+2h_{\Yboxdim{3pt}(\yng(1),\yng(1))}-2h_{\Yboxdim{3pt}(\yng(1),0)}}}.
\fe
By taking the derivative $\partial_{z_1}$ and taking the large $N$ limit, we obtain
\ie
&\vev{\partial\omega_1(z_1)\phi_1(z_2)(\overline{\yng(2)},\overline{\yng(1)})(z_3)}={1\over \sqrt{2}}{1\over |z_{23}|^{2(1-\lambda)}}\left({\lambda\over N}{1\over z_{12}}-{\lambda+\lambda^2\over N}{1\over z_{12}}\right),
\\
&\vev{\partial\omega_1(z_1)\phi_1(z_2)(\overline{\yng(1,1)},\overline{\yng(1)})(z_3)}={1\over \sqrt{2}}{1\over |z_{23}|^{2(1-\lambda)}}\left(-{\lambda\over N}{1\over z_{12}}+{\lambda-\lambda^2\over N}{1\over z_{12}}\right).
\fe
Taking the difference of these two three-point functions, we obtain
\ie
&\vev{  j^{(1)}_1(z_1)\phi_1(z_2)\bar\phi_2(z_3)}={1\over \sqrt{N}}{1\over |z_{23}|^{2(1+\lambda)}}\left({1\over z_{12}}-{1\over z_{13}}\right).
\fe
In a similar way, we also compute the three-point functions
\ie
&\vev{  j^{(1)}_1(z_1)\phi_2(z_2)\bar\phi_3(z_3)}=\vev{  j^{(1)}_2(z_1)\phi_1(z_2)\bar\phi_3(z_3)}={1\over \sqrt{N}}{1\over |z_{23}|^{2(1+\lambda)}}\left({1\over z_{12}}-{1\over z_{13}}\right),
\fe
and also
\ie
&\vev{  j^{(1)}_1(z_1)\tilde\phi_1(z_2)\bar{\tilde\phi}_2(z_3)}=\vev{  j^{(1)}_1(z_1)\tilde\phi_2(z_2)\bar{\tilde\phi}_3(z_3)}=\vev{  j^{(1)}_2(z_1)\tilde\phi_1(z_2)\bar{\tilde\phi}_3(z_3)}
\\
&=-{1\over \sqrt{N}}{1\over |z_{23}|^{2(1-\lambda)}}\left({1\over z_{12}}-{1\over z_{13}}\right).
\fe
We postulate the general form of these kind of three-point functions to be
\ie
&\vev{  j^{(1)}_n(z_1)\phi_m(z_2)\bar\phi_{n+m}(z_3)}={1\over \sqrt{N}}{1\over |z_{23}|^{2(1+\lambda)}}\left({1\over z_{12}}-{1\over z_{13}}\right),
\\
&\vev{  j^{(1)}_n(z_1)\tilde\phi_m(z_2)\bar{\tilde\phi}_{n+m}(z_3)}=-{1\over \sqrt{N}}{1\over |z_{23}|^{2(1-\lambda)}}\left({1\over z_{12}}-{1\over z_{13}}\right).
\fe

\end{document}